\documentclass[usenatbib]{mnras}
\usepackage{multirow}
\usepackage{graphicx}
\usepackage[space]{grffile}
\usepackage{latexsym}
\usepackage{amsfonts,amsmath,amssymb}
\usepackage{url}
\usepackage[utf8]{inputenc}
\usepackage{hyperref}
\hypersetup{colorlinks=false,pdfborder={0 0 0}}
\usepackage{textcomp}
\usepackage{longtable}
\usepackage{natbib}

\title[UKIRT followup of Y0 dwarfs]{Parallaxes and Infrared Photometry of three Y0 dwarfs}

\author[R. L. Smart et al.]
{R. L. Smart$^{1,6}$\thanks{E-mail: smart@oato.inaf.it, Leverhulme Visiting Professor},
D\'aniel Apai$^{2,3}$,
J. Davy Kirkpatrick$^4$,
S. K. Leggett$^5$,
F. Marocco$^6$,
\newauthor
Jane E. Morrison$^2$,
H. R. A. Jones$^6$,
D. Pinfield$^6$,
P. Tremblin$^7$
and D.S. Amundsen$^8$.
\\
$^{1}$Istituto Nazionale di Astrofisica, Osservatorio Astrofisico di Torino, Strada Osservatorio 20, 10025 Pino Torinese, Italy\\
$^{2}$Steward Observatory, 933 N. Cherry Avenue, University of Arizona, Tucson, AZ 85721, USA\\ 
$^{3}$Lunar and Planetary Laboratory, 1629 E. University Boulevard, University of Arizona, Tucson, AZ 85721, USA\\ 
$^{4}$Infrared Processing and Analysis Center, MS 100-22, California Institute of Technology, Pasadena, CA 91125, USA\\
$^{5}$Gemini Observatory, 670 N. A'ohoku Place, Hilo, HI 96720, USA\\
$^{6}$School of Physics, Astronomy and Mathematics, University of Hertfordshire, College Lane, Hatfield AL10 9AB, UK\\
$^{7}$Maison de la Simulation, CEA-CNRS-INRIA-UPS-UVSQ, USR 3441, Centre
d’\'etude de Saclay, F-91191 Gif-Sur-Yvette, France\\
$^8$Astrophysics Group, University of Exeter, EX4 4QL Exeter, UK
}

\begin{document}

\date{Accepted . Received ; in original form 2016 June 19}


\maketitle

\begin{abstract}

  We have followed up the three Y0 dwarfs WISEPA
  J041022.71+150248.5, WISEPA J173835.53+273258.9 and WISEPC
  J205628.90+145953.3 using the UKIRT/WFCAM
  telescope/instruments. We find parallaxes that are more
  consistent and accurate than previously published values.  We
  estimate absolute magnitudes in photometric pass-bands from $Y$
  to $W3$ and find them to be consistent between the three Y0
  dwarfs indicating the inherent cosmic absolute magnitude spread
  of these objects is small. We examine the MKO $J$ magnitudes
  over the four year time line and find small but significant
  monotonic variations.  Finally we estimate physical parameters
  from a comparison of spectra and parallax to equilibrium and
  non-equilibrium models finding values consistent with solar
  metallicity, an effective temperature of 450-475\,K and log~g
  of 4.0-4.5.

\end{abstract}
\begin{keywords}
astrometry -- parallaxes -- brown dwarfs -- techniques:spectroscopic
\end{keywords}

\section{Introduction}

Y dwarfs represent the coolest collapsed objects outside the solar
system known to date. They exhibit strong methane absorption where the
Wide-field Infrared Survey Explorer mission \cite[WISE;
][]{2010AJ....140.1868W} $W1$ 3.4\,$\mu$m filter is centered and emit
about half their energy in the WISE $W2$ 4.6\,$\mu$m pass band
\citep{2011ApJ...726...30M}. This makes the $W1-W2$ color very distinct
for these objects and most of the known Y dwarfs have been discovered
following their identification as colour-selected candidates in the
WISE data \cite[e.g.][]{2011ApJ...743...50C}.


The temperatures of the Y0 subclass dwarfs are believed to be around 400\,K
and their masses to be between 5-30 $M_{Jup}$ \citep{2011ApJ...743...50C},
overlapping in physical parameter space with many exoplanets, so they can be
used as surrogates to understand the atmospheric processes of exoplanets. The
older examples will hold the chemical imprint of the early Galaxy and the
distribution in age may help map out the evolution of formation mechanisms
over the Galaxy's lifetime.

In this contribution we discuss the three objects WISEPA J041022.71+150248.5,
WISEPA J173835.53+273258.9 and WISEPC J205628.90+145953.3 which will refer to as
0410, 1738 and 2056 respectively.  These were all originally presented in
\cite{2011ApJ...743...50C} and classified as spectral types Y0.  First we
discuss the astrometry, then the photometry and finally we combine these
observations with published spectra and models to estimate physical
parameters.
%
%
\section{Astrometric Analysis}

\begin{table*}
\caption{\label{parallaxes}Parallaxes and proper motions for UKIRT Y0 targets.}
\centering
\begin{tabular}{lrrrrrrrrrrr}
\hline\hline
Short   & ~~~RA,  ~~~  ~~~ Dec ~~~  &  Epoch    &  Absolute $\pi$  &  $\mu_{\alpha}$~~~~ & $\mu_{\delta}$~~~~  & COR    &$N_*,N_e$&$\Delta$T &$V_{tan}$\\
Name    & (h:m:s),~~   ($^\circ$:':'')~~& (yr)~~ &  (mas) ~~        &   (mas/yr)   ~    & (mas/yr)   ~       &(mas)  &         & (yr)     & (km/s)\\
\hline 
0410   &  4:10:23.0, +15:02:37.8 & 2014.9551 &   144.3 $\pm$  9.9 &   956.8 $\pm$  5.6 & -2221.2 $\pm$  5.5 &   0.97 &  99, 19 &   4.36 &    79.4 $\pm$  5.4\\
1738   & 17:38:35.6, +27:32:57.8 & 2013.2584 &   128.5 $\pm$  6.3 &   345.0 $\pm$  5.7 &  -340.1 $\pm$  5.1 &   0.63 & 293, 18 &   4.53 &    17.9 $\pm$  0.9\\
2056   & 20:56:29.0, +14:59:54.6 & 2012.8274 &   148.9 $\pm$  8.2 &   826.4 $\pm$  5.5 &   530.7 $\pm$  8.5 &   0.67 & 452, 18 &   4.54 &    31.3 $\pm$  1.7\\
                                                                                                                                                   
\hline
\end{tabular} \\
COR = correction to absolute parallax, $N_*$ = number of
reference stars, $N_e$ = number of epochs, $\Delta$T = epoch range, $V_{tan}$= tangential velocity.
\end{table*}

\subsection{Observational Data}

The astrometric observations were all made on the UKIRT 3.8~m telescope using
the WFCAM imager, which was the combination used to produce the UKIDSS surveys
\citep[e.g.][]{2007MNRAS.375..213W}.  All observations are carried out in
queue override mode, allowing us to be very flexible in the scheduling,
maximising the parallax factor and observing close to meridian passage.  The
first observations of these objects were made in September 2011 following a
Director Discretionary Time request (U/11B/D1). During the 2012A
semester they were included as part of the UKIRT ultra cool dwarf parallax
program described in \citet[][]{2010AA...511A..30S} and
\citet[][hereafter MSJ10]{2010AA...524A..38M}. In 2014 via a request to the
University of Arizona (U/14B/UA15) we obtained further observations. The
results published here are based on observations from September 2011 to April
2016. The basic procedures for observing, image treatment and parallax
determination follow those described in MSJ10.

One of the most important aspects in the determination of small field
parallaxes is stability of the focal plane and repetition of the
observational procedure. This is particularly true for the large off
axis detectors of the WFCAM instrument. In the MSJ10 program we required
that the targets are observed in the same physical position on the
focal plane as the discovery image in the UKIDSS survey. 
As these Y0 dwarfs were not in the UKIDSS survey we had the ability to place
the target on any of the 4 chips. For
1738 and 2056 they were placed in the most central quadrant -
with respect to the optical axis - of chip 3. This region being close to the
optical axis is astrometrically ``quiet''.  For 0410, in an attempt
to also include the T6 dwarf WISEP J041054.48+141131.6 in chip 1, we
placed it in the top outside quadrant of chip 4. Unfortunately the T6
is not in the chip 1 as we hoped, but once the first image was taken
we kept the same relative position.

For each target we only used reference stars within a limited radius.  The
size of the radius is a compromise between limiting the number of reference
stars and having a large astrometrically complex area to transform. For all
targets we choose 2 arcminutes which provided over 50
reference stars. There is a factor of 4 difference between the number of
reference stars for 2056 vs 0410 (see $N_*$ in Table~\ref{parallaxes}) but
50 was still considered sufficient to astrometrically model such a small area.

As these objects are fainter in the $J$ band than the other targets 
in the MSJ10 
program we increased the exposure time following a 5 jitter
(dithered) 3.2'' cross pattern, and at each jitter position we made 4
exposures in 2$\times$2 micro-stepped positions of 1.5 pixels, where each
exposure consists of 4 co-added 10~second images. The total exposure time is
therefore $5\times 4 \times 4 \times 10 = 800$~seconds. In average conditions
this provides a signal-to-noise of 50 at MKO $J=19.5$.

All observations are reduced using the standard WFCAM Cambridge Astronomical
Survey Unit (CASU) pipeline. We transformed all frames to a base frame using a
simple six constant linear astrometric fit. We then removed any frames that
have an average reference star error larger than the mean error for all frames
plus three standard deviations about that mean in either coordinate, or, have
less than 12 stars in common with the base frame. Each observation has a
quoted positional error from the UKIRT pipeline based the profile fitting
program, and the errors in the transformation parameters. However, there
remains a systematic contribution to the error that changes from night
to night.  For this reason, when fitting for the astrometric parameters to the
individual observations on the combined frames we
treat each observation with equal weight and then calculated the final error on
the target parameters from the co-variance matrix of the solution scaled by
the error of unit weight. This fit is also iterated removing any observations
where the combined residual in the two coordinates is greater than three times
the sigma of the whole solution.

The solutions were tested for robustness using bootstrap-like testing
where we iterate through the sequence selecting different frames as
the base frame thus making many solutions that incorporate slightly
different sets of reference stars and starting from different
dates. We select all solutions with: (i) a parallax within one sigma
of the median solution; (ii) the number of included observations in
the top 10\%, and (iii) at least 20 reference stars in common to all
frames. From this subset, for this publication, we have selected
the one with the smallest error. More than 90\% of the solutions are
within one sigma of the published solution. 

To these relative parallaxes we add a correction (COR in Table~\ref{parallaxes}) to find the astrophysically useful absolute parallaxes. The
COR is estimated from the average magnitude of the reference stars and the
model of Mendez \& van Altena (1996\nocite{men96}) transformed into the J
band.

\begin{figure}
\centering
\includegraphics[width=7.4cm]{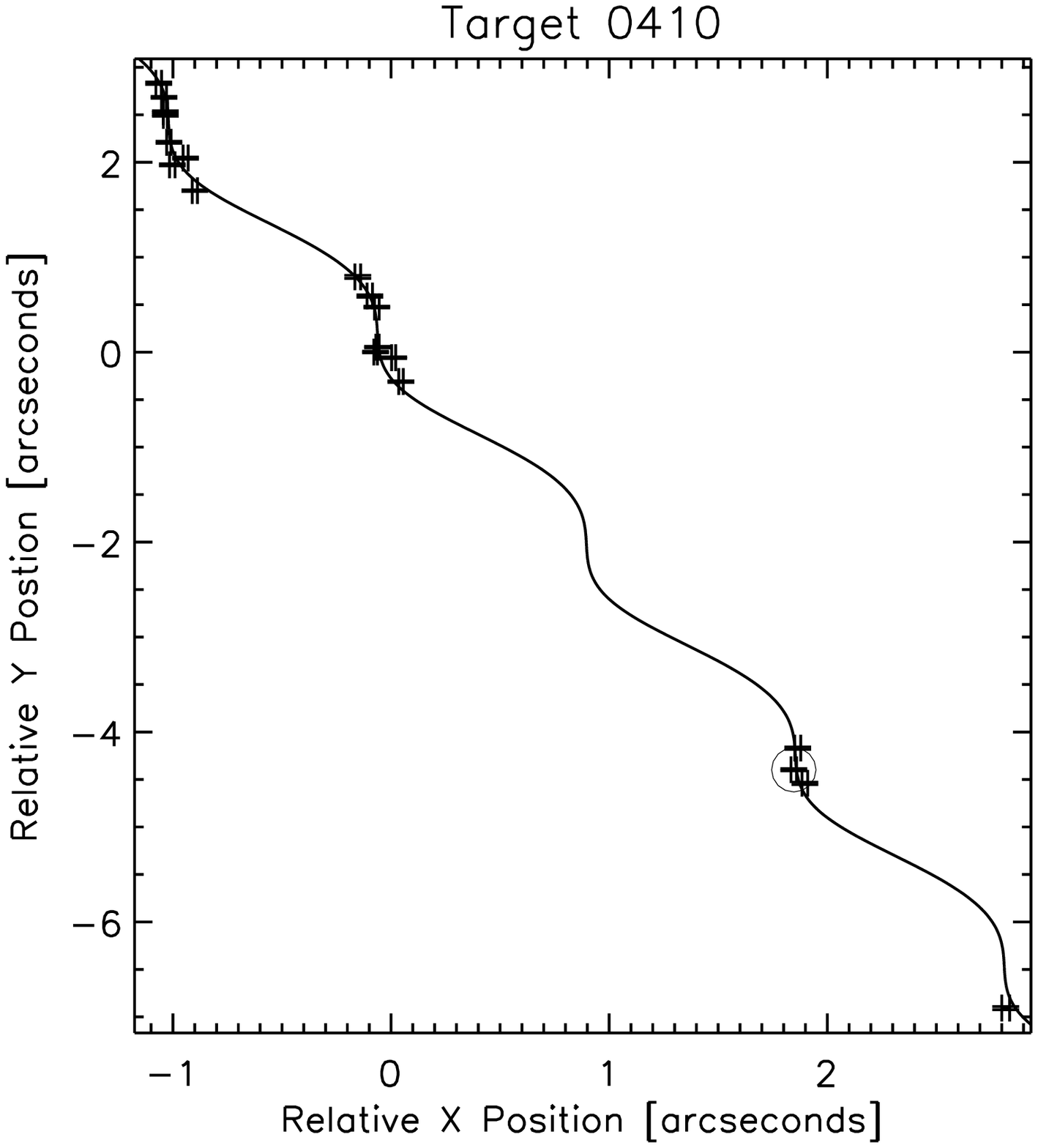}
\includegraphics[width=7.4cm]{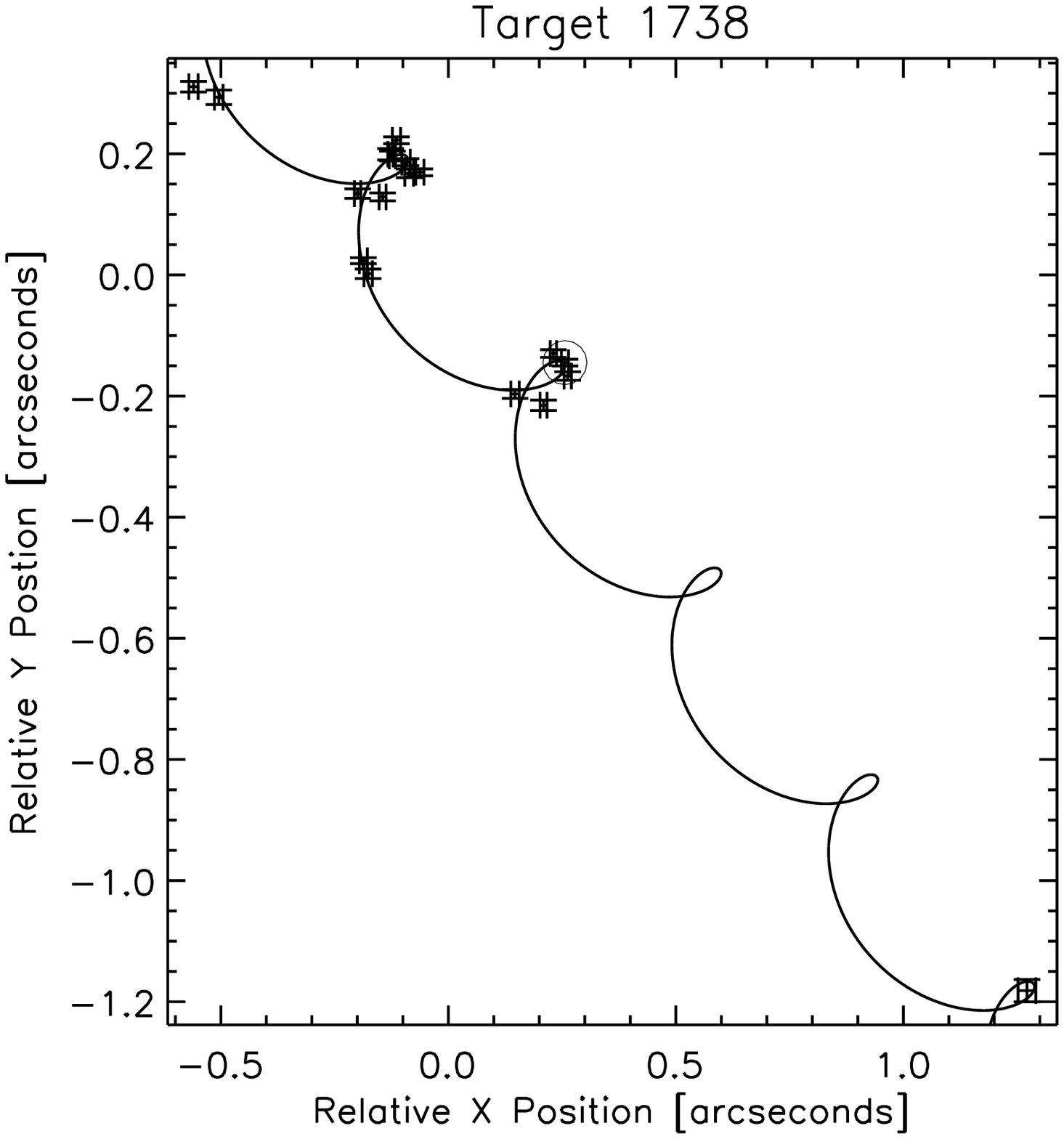}
\includegraphics[width=7.4cm]{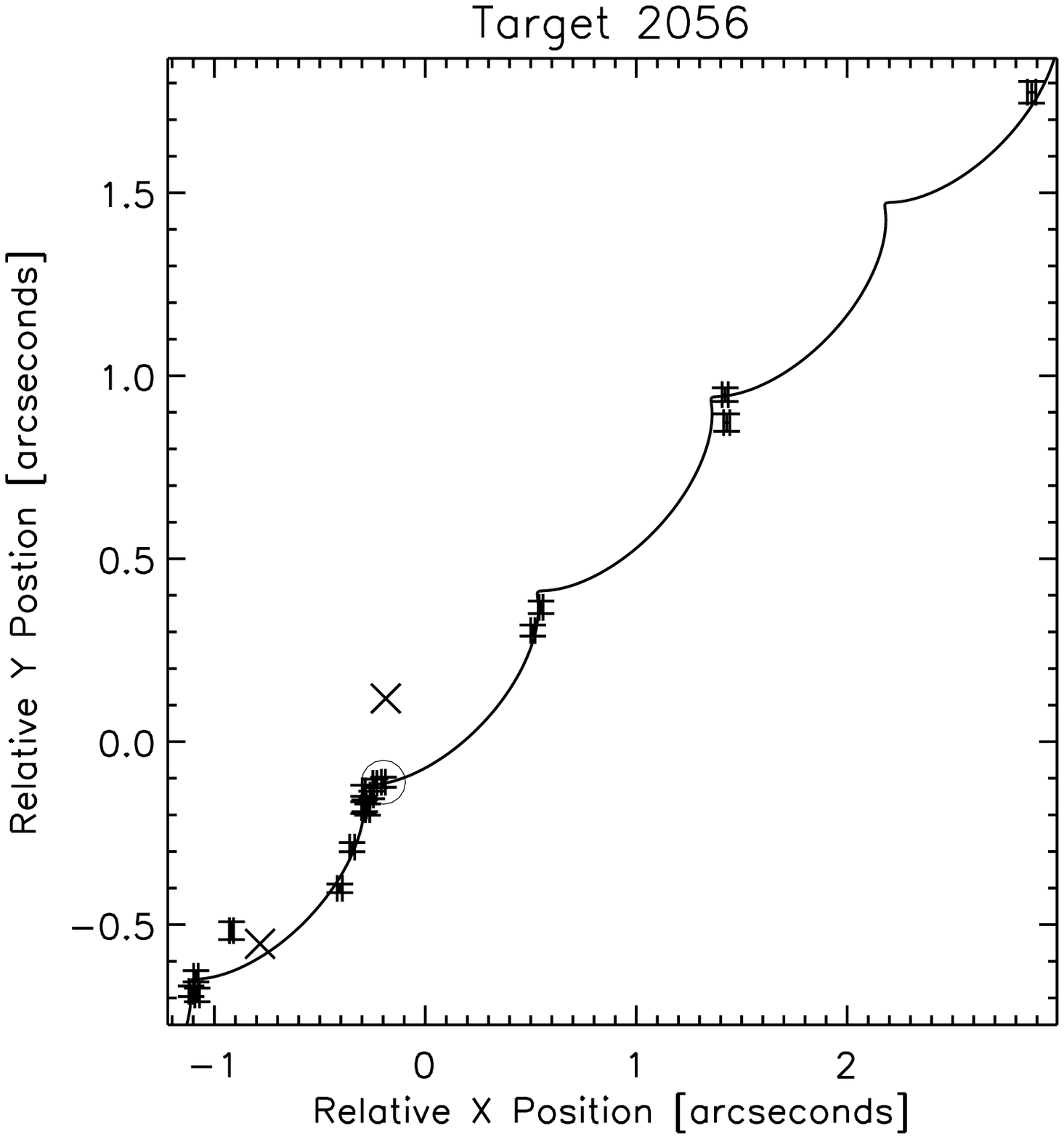}
\caption{Observations and solution for 0410, 1738 and 2056 from top to bottom
  panels respectively. The base frame selected is indicated by a circle and any
  observations rejected are indicated by X. The error bars are the formal
  centroiding errors from the CASU pipeline in both coordinates. }
\label{TD04105_par_pap4.eps}
\end{figure}

\subsection{Astrometric Parameters}

In Table~\ref{parallaxes} we report the derived absolute parallaxes, proper
motions and details of the solutions for the UKIRT sequences.  In Fig.
\ref{TD04105_par_pap4.eps} we plot the observations and the predicted movement
of the targets from our parameters.    The result for 0410 is
lower precision than 1738 and 2056 probably due to the higher proper motion,
since any error in our estimation of this motion will propagate into the
parallax estimate.

\subsection{Comparison to published values}

Parallaxes for these objects have been measured by three teams
\citet{2013ApJ...762..119M}, \citet{2014ApJ...783...68B}, and
\citet[][hereafter D\&K13]{2013Sci...341.1492D}. The Marsh et al. and and
Beichman et al. works used combinations of WISE $W2$, WIRC $J$, NEWFIRM,
Spitzer channel 2, and HST $J$ observations with a significant overlap of the
two sets of observations. The Marsh results for these targets were based on
7-9 observations while those of Beichman, coming later, were based on
14-16. The results for D\&K13 were based on 5 Spitzer channel 1 observations.
In \citet{2011ApJS..197...19K} they also provide parallactic distances but
they were preliminary estimates from the \citet{2013ApJ...762..119M} work so
we have only reported the latter values. Considering all the published targets
there are 9 Y0 dwarfs with more than one estimated parallax including the
three objects under study here.  In Table~\ref{pitable} we have reported all
results in common between the three cited works and the results from this
contribution.

\begin{table}
\caption{\label{pitable}Comparison of parallaxes and proper motions for
  Y0 dwarfs in different programs.}
\begin{center}
\begin{tabular}{lllll}
\hline\hline
Short   &  $\mu_{\alpha}$~~~~ & $\mu_{\delta}$~~~~ &  Absolute $\pi$ & Ref\\
~~Name  &  (mas/yr)   ~    & (mas/yr)   ~      &   (mas)          &    \\
\hline 
0254 &  2588$\pm$27  &   273$\pm$27  &  135$\pm$15  &  2 \\
0254 &  2578$\pm$42  &   309$\pm$50  &  185$\pm$42  &  3\\
\hline 
0410 &   966$\pm$13  & -2218$\pm$13  &  160$\pm$9   &  1 \\
0410 &   958$\pm$37  & -2229$\pm$29  &  132$\pm$15  &  2    \\
0410 &   974$\pm$79  & -2144$\pm$72  &  233$\pm$56  &  3    \\
0410 &   956$\pm$06  & -2223$\pm$6   &  144$\pm$10  &  4 \\
\hline   
1405 & -2263$\pm$47  &   288$\pm$41  &  129$\pm$19  &  2    \\
1405 & -2297$\pm$96  &   212$\pm$137 &  133$\pm$81  &  3    \\
\hline  
1541 &  -857$\pm$12  &  -087$\pm$13  &  176$\pm$9   &  1 \\
1541 &  -870$\pm$130 &  -013$\pm$58  &   74$\pm$31  &  2    \\
1541 &  -983$\pm$111 &  -276$\pm$116 &  -21$\pm$94  &  3    \\
\hline    
1738 &   317$\pm$09  &  -321$\pm$11  &  128$\pm$10  &  1 \\
1738 &   292$\pm$63  &  -396$\pm$22  &  102$\pm$18  &  2    \\
1738 &   348$\pm$71  &  -354$\pm$55  &   66$\pm$50  &  3    \\
1738 &   346$\pm$6   &  -338$\pm$5   &  129$\pm$6   &  4    \\
\hline 
1741 &  -509$\pm$35  & -1463$\pm$32  &  180$\pm$15  &  2    \\
1741 &  -495$\pm$11  & -1472$\pm$13  &  176$\pm$26  &  3    \\
\hline 
1804 &  -269$\pm$10  &   035$\pm$11  &   80$\pm$10  &  1 \\
1804 &  -242$\pm$26  &   017$\pm$22  &   60$\pm$11  &  2    \\
\hline
1828 &  1024$\pm$7   &   174$\pm$6   &  106$\pm$7   &  1 \\
1828 &  1020$\pm$15  &   173$\pm$16  &   70$\pm$14  &  2    \\
\hline
2056 &   812$\pm$9   &   534$\pm$8   &  140$\pm$9   &  1 \\
2056 &   761$\pm$46  &   500$\pm$21  &  144$\pm$23  &  2    \\
2056 &   881$\pm$57  &   544$\pm$42  &  144$\pm$44  &  3    \\
2056 &   828$\pm$6   &   532$\pm$8   &  149$\pm$8   &  4 \\
\hline
\end{tabular}
\end{center}
References: 1:\citet{2014ApJ...783...68B},
2:\citet{2013Sci...341.1492D}, 3:\citet{2013ApJ...762..119M} and 4:
this work. Note D\&K13 only publish relative parallaxes which is what we have
reported here.
The targets are 0254: WISE J025409.45+022359.1, 0410: WISEPA J041022.71+150248.5
1405: WISE J140518.40+553421.5,
1541: WISE J154151.65--225025.2,
1738: WISE J173835.52+273258.9,
1741: WISE J174124.26+255319.5,
1804: WISE J180435.40+311706.1,
1828: WISE J182831.08+265037.8,
2056: WISE J205628.90+145953.3.
\end{table}

\begin{figure}
\centering
\includegraphics[width=9cm]{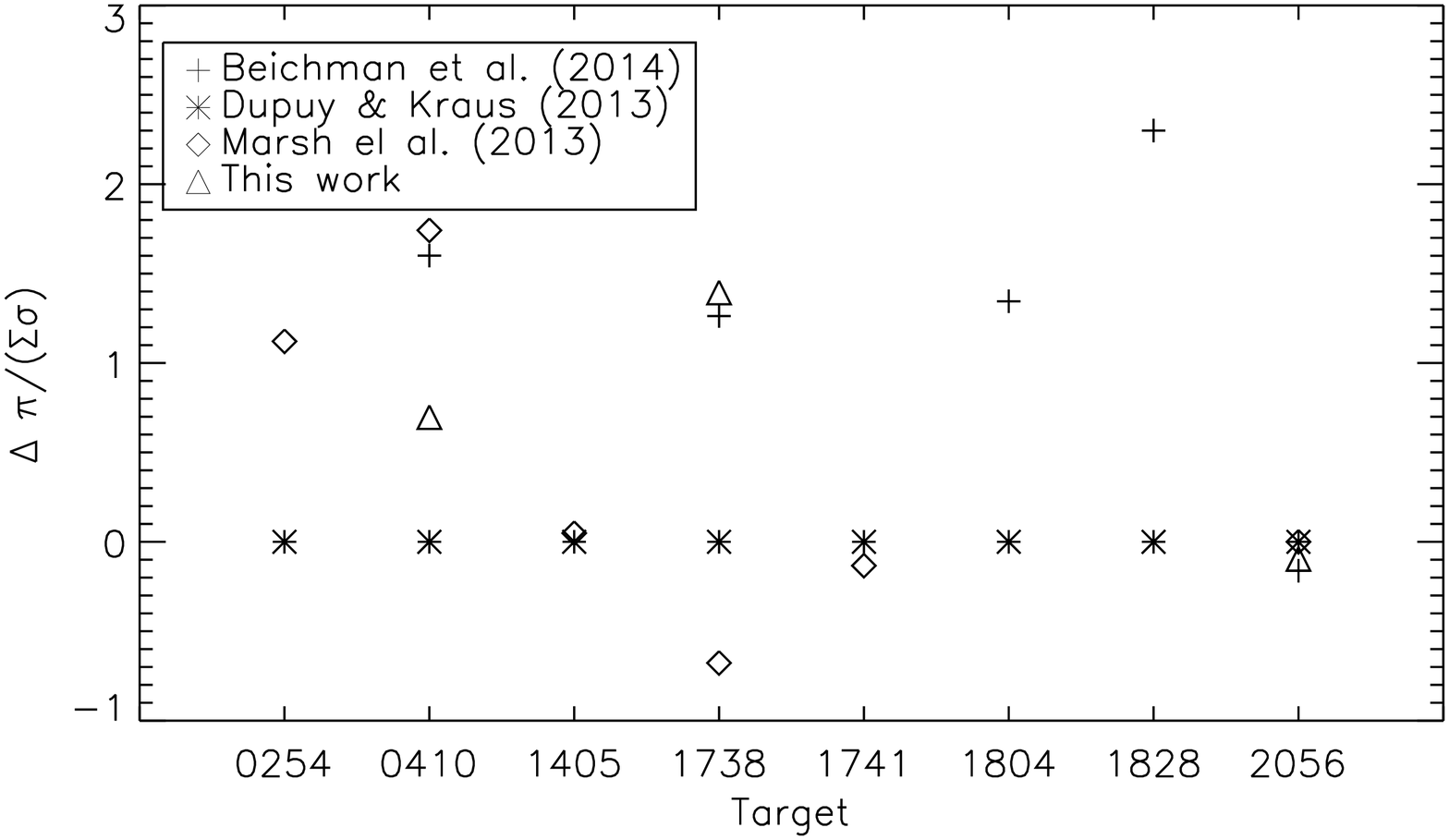}
\caption{Parallaxes differences compared to the D\&K13 values as multiples of the combined
  standard deviation for each solution. The D\&K13 value is used as the reference as they have published 
  a parallaxes for all of the Y0 dwarfs considered here. References and target
  abbreviations as in Table~\ref{pitable}. So for example the Marsh
  value of 0254 is one combined sigma larger than the D\&K13 value.}
\label{comparison_forpap.eps}
\end{figure}

In Fig.~\ref{comparison_forpap.eps} we plot the differences of the parallax
values for the 9 targets with respect to the D\&K13 value divided by the error
of the two estimates combined in quadrature. The D\&K13 values are relative
but the difference between relative and absolute parallax is negligible
compared to the random errors. The 1541 and 1738 Marsh et al. values are very
different from the other values for these targets and are also significantly
lower than the predicted spectroscopic parallax so we assume these are
compromised. Given the short time baseline and mixed observational sources for
the Marsh et al. work it is to be expected that in some solutions the proper
motion and parallax were not disentangled correctly.  Apart from the two low
Marsh et al. values all the other D\&K13 estimates appear as underestimates,
on average by one combined sigma.

D\&K13 publish relative parallaxes not absolute ones because they felt the
correction was negligible. The corrections we have applied are less than
1\,mas and since the average reference star is fainter in the Spitzer fields
we would expect the D\&K13 corrections to be even smaller.  A correction will
reduce the difference in the right direction but we agree with the authors
that it cannot be the main cause of the observed difference.  We also note in
\cite{2014ApJ...796...39T} they also find the D\&K13 parallax estimates are
low for other targets.  This comparison to the D\&K13 values indicates the
most reliable results are those of Beichman et al. The difference between the
results published here and those of Beichman et al. are all within one
sigma. We consider this a confirmation of our procedures, parallax estimates
and, importantly, error estimates. 

\subsection{Search for common proper motion objects and moving group membership}

We searched for common proper motion companions to our 3 Y dwarfs
within the Hipparcos Main Catalogue, the Gliese Catalogue of
Nearby Stars, the Tycho-2 catalogue, and the Fourth U.S. Naval
Observatory CCD Astrograph Catalogue. We looked for objects with
differences in both proper motion components $<3\sigma$ and a
maximum projected separation $<100,000$~AU. The search returned
no matches.

We used the BANYAN II online
tool\footnote{\url{http://www.astro.umontreal.ca/\~gagne/banyanII.php}}
\citep{2013ApJ...762...88M,2014ApJ...783..121G} to assess the membership of
our targets to nearby moving groups. None of the targets have significant
probability of belonging to any of the moving groups. However, for 2056, we
obtained a probability of $44\%$ to be a old field member and $56\%$ to be a
young field member, suggesting that this object might pertain to a slightly
younger population. The tangential velocity, listed in the last column of
Table~\ref{parallaxes}, of 0410 is significantly larger than both 1738 and
2056 suggesting it might be old but it does not exceed the $V_{tan} >
100$\,km\,s$^{-1}$ criteria adopted for ultracool dwarfs belonging to either
the Galactic thick disk or halo \citep{2009AJ....137....1F}.

We also used LACEwING\footnote{\url{https://github.com/ariedel/lacewing}} to
access membership assuming the three targets are field objects - e.g. that we
have no evidence of youth. Again none of the targets show a significant
probability of being in a moving group though 2056 did have a 30\% probability
of belonging to the Argus group. We conclude that these objects are not
members of any know moving groups and are probably just local field members.

\section{Photometric Analysis} 
\subsection{Photometric Analysis of WFCAM data}
The CASU pipeline estimates the MKO $J$ magnitude of our targets using
~100-200 calibrating stars from the Two Micron All Sky Survey \cite[][,
  hereafter 2MASS]{2006AJ....131.1163S} as described in
\cite{2009MNRAS.394..675H}. The systematic errors of the calibration from
2MASS stars is estimated to be better than 1.5\% (Hodgkin et al. 2009). The
random error is calculated in the pipeline procedures. Following the
recommendation in \cite{2006MNRAS.372.1227D} we have adopted the {\bf aperMag3
} parameter as the best estimate of the total magnitude for point sources.

\begin{figure}
\centering
\includegraphics[width=9cm]{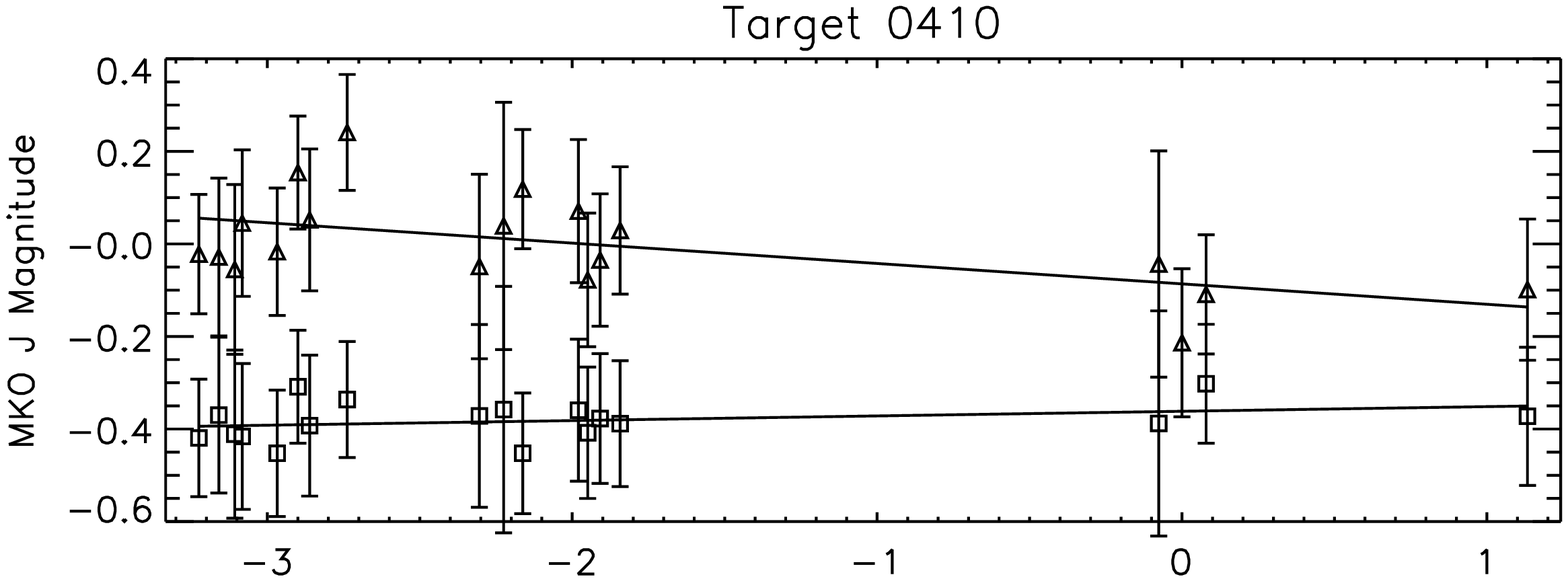}
\includegraphics[width=9cm]{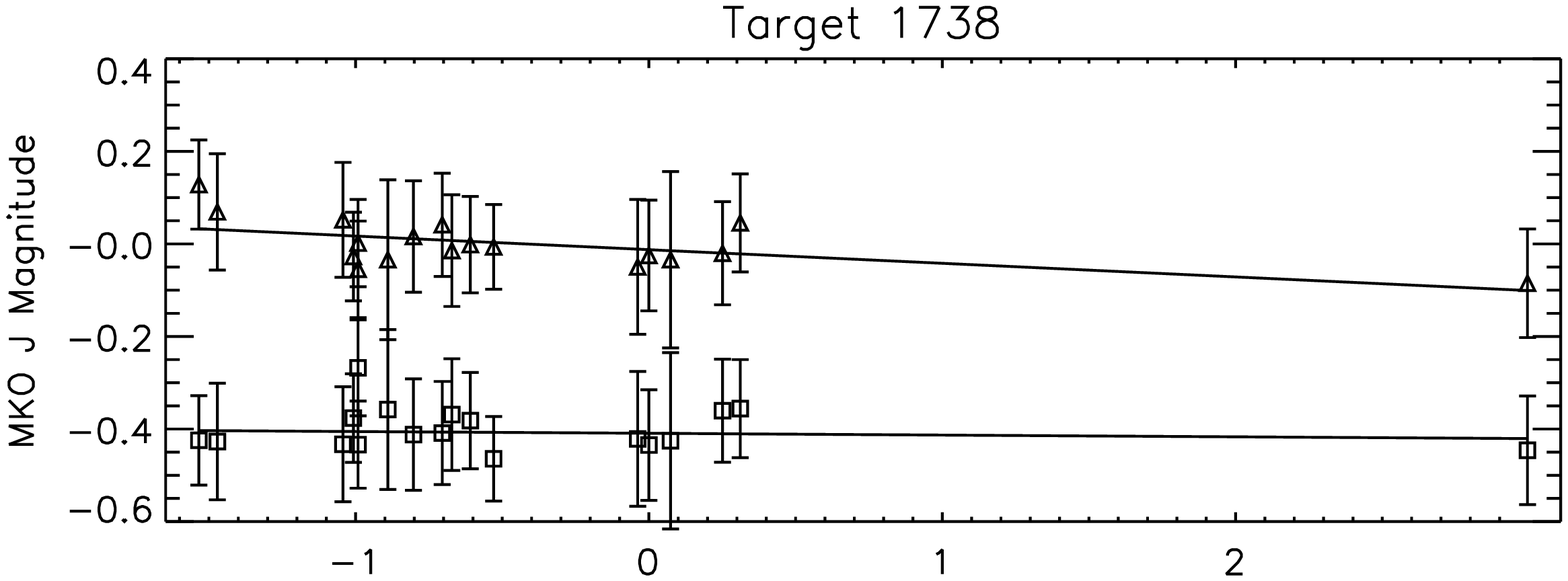}
\includegraphics[width=9cm]{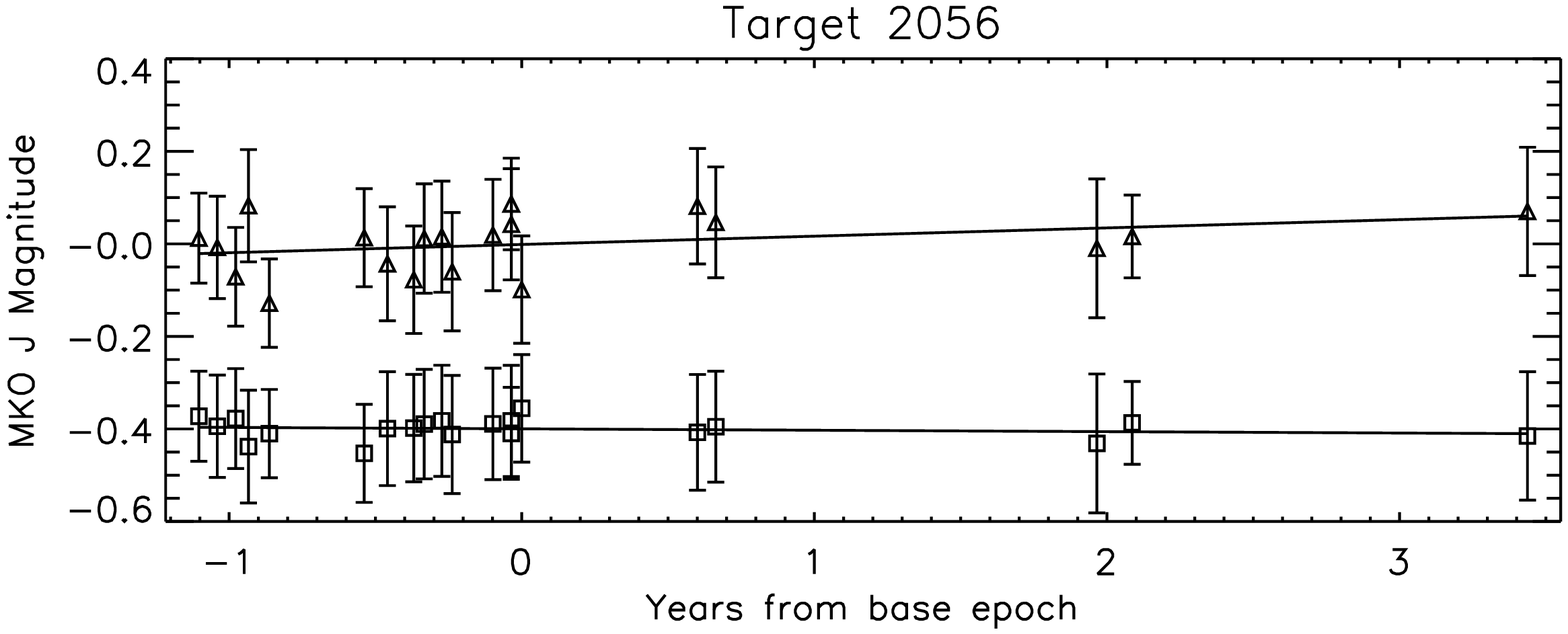}
\caption{ MKO $J$ magnitude variations as a function of time for 0410, 1738
  and 2056 respectively. In each plot we have included an anonymous object
  that is nearby in position and magnitude and plotted it's magnitude
  variation offset by -0.4\,mag.}
\label{TD20565.magplot0}
\end{figure}

In Fig.~\ref{TD20565.magplot0} we plot the variation of the $J$ magnitude for
the parallax program observations with respect to the mid-epoch. In
Table~\ref{spitzermagstable} we report the number of observations, the
magnitude, standard deviation, the error of the mean and the slope of the best
fit straight line. The long term changes of 0410 and 1738 do appear to be
significant, indicating a slow dimming over the period observed, the number of
observations is, however, very low and quite noisy. Further observations of
these targets are needed to confirm the observed trends.

We adopt the mean magnitude from Table~\ref{spitzermagstable} as the best
estimate of the MKO magnitude for these targets. To provide a conservative
estimate of the error of the mean we simply add 0.02 magnitudes to account for
the 1.5\% systematic error estimates from Hodgkin et al. (2009).


\begin{table}
\caption{\label{spitzermagstable}Mean MKO $J$ magnitudes and variations with time.}
\begin{center}
\begin{tabular}{lrrrrr}
\hline\hline
Target &  N &  $< J >$ & $\sigma$ & $\sigma_{mean}$ &     Slope \\
       &    &  mag     &  mag     &  mag           & mag/yr \\
\hline 
0410 &  19 & 19.137 &  0.103 &  0.031 & -0.044 $\pm$   0.016     \\
Anon &     & 18.789 &  0.219 &  0.054 &  0.010 $\pm$   0.038    \\
1738 &  18 & 19.539 &  0.051 &  0.023 & -0.029 $\pm$   0.010     \\
Anon &     & 19.048 &  0.068 &  0.026 & -0.004 $\pm$   0.058     \\
2056 &  20 & 19.237 &  0.062 &  0.024 &  0.018 $\pm$   0.012     \\
Anon &     & 18.593 &  0.005 &  0.020 & -0.003 $\pm$   0.005     \\
\hline
\end{tabular}
\end{center}
The ``Anon'' entry for each target is the straight line fit to the
  anonymous field star in each sequence. The slope is always smaller and
  within one sigma zero.
\end{table}

\subsection{Literature Photometry}
There are three published MKO $J$ and $H$ values from
\cite{2015ApJ...799...37L}, \cite{2015ApJ...804...92S} and
\cite{2012ApJ...753..156K} for these targets as reported in
Table~\ref{MKOphotometry} along with the estimates found in the last
section. The differences are significant, reaching 1 magnitude for the H band
of target 0410. The Leggett et al. (2015) values are consistently fainter than
all these values and this was discussed in that paper. Since the number of
estimates is too small to allow any meaningful statistical tests we simply
calculate a weighted mean of all values as reported in
Table~\ref{MKOphotometry} use this in our further analysis.

\begin{table}
\caption{\label{MKOphotometry}Published MKO $J$ and $H$ magnitudes and
  weighted means.}  \centering
\begin{tabular}{lrrrrr}
\hline\hline
Target &  MKO $J$, $\sigma$  &  MKO $H$,$\sigma$     &    Source \\
\hline  
0410 &   19.44, 0.03 &  20.02, 0.05 & Leggett et al. 2015  \\     
    &   19.33, 0.02 &  19.90, 0.04 & Schneider et al. 2015 \\    
    &   19.24, 0.05 &  19.05, 0.09 & Kirkpatrick et al. 2012 \\  
    &   19.23, 0.05 &     ...      & Mean from Table~\ref{spitzermagstable} \\	 	    
    &   19.34, 0.02 &  19.85, 0.03 & Weighted mean \\            
\hline  
1738 &   19.63, 0.05 &  20.24, 0.08 & Leggett et al. 2015  \\    
     &   19.55, 0.02 &  20.25, 0.03 & Schneider et al. 2015 \\   
     &   19.51, 0.08 &  20.39, 0.33 & Kirkpatrick et al. 2012 \\ 
     &   19.52, 0.03 &    ...       & Mean from Table~\ref{spitzermagstable} \\	 	  
     &   19.55, 0.02 &  20.25, 0.03 & Weighted mean \\           
\hline  \\
2056 &   19.43, 0.04 &  19.96, 0.04 & Leggett et al. 2015  \\    
     &   19.13, 0.02 &  19.64, 0.03 & Schneider et al. 2015 \\   
     &   19.23, 0.13 &  19.62, 0.31 & Kirkpatrick et al. 2012 \\ 
     &   19.23, 0.03 &   ...        & Mean from Table~\ref{spitzermagstable}  \\	 	  
     &   19.20, 0.02 &  19.75, 0.02 & Weighted mean \\           
\hline
\end{tabular}
\end{table}

In Table~\ref{allphotometry} along with the weighted mean $J$ and $H$
magnitudes we report all published photometry from other bands.  From the
apparent magnitudes we estimate absolute magnitudes assuming a distance given
by the parallax in Table~\ref{parallaxes}. The error on the distance is the
largest contributor to the error in the absolute magnitude.  In the last line
of Table~\ref{allphotometry} we include a weighted mean absolute magnitude and
the error of that mean.

\begin{figure}
\centering
\includegraphics[width=9cm]{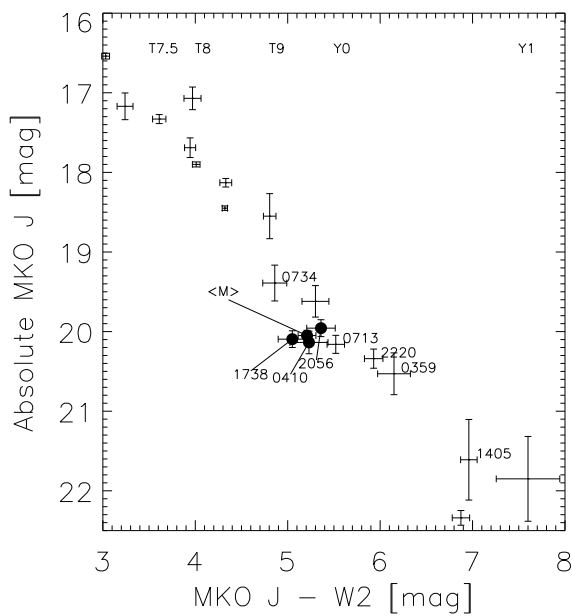}
\caption{The absolute MKO $J$ magnitude versus the MKO $J$ - W2 color for
  published $>$T7 objects from Leggett et al. (2015) along with the objects
  presented here.  The spectral types at the top of the graph are provided
  just as indicative ranges, the labelled objects are the Y0 dwarfs 0359 WISE
  J035934.06--540154.6, 0713: WISE J071322.55--291751.9, 0734: WISE
  J073444.02--715744.0, 1405: WISEP J140518.40+553421.5 and 2220: WISE
  J222055.31--362817.4. The $<$M$>$ point is the weighted mean absolute
  magnitude as found in Table~\ref{allphotometry}.  }
\label{wise_t6_ownphot_tinneypi2}
\end{figure}

In Fig.~\ref{wise_t6_ownphot_tinneypi2} we plot the absolute MKO $J$
magnitudes vs MKO $J -$ W2 color for various $>$T7 objects with distances and
magnitudes taken from \cite{2015ApJ...799...37L}. The three targets presented
here are labelled, as are the other Y0 objects in the sample. The weighted
mean absolute $J$ and $W2$ magnitudes, 20.05$\pm$0.07 and 14.84$\pm$0.07, from
Table~\ref{allphotometry} can be directly compared to the median absolute
magnitudes based on 11 Y0 dwarfs with measured distances in
\cite{2014ApJ...796...39T} of 20.32$\pm$1.25 and 14.65$\pm$0.35. The
difference is large but within 1 sigma and as seen from
Fig.~\ref{wise_t6_ownphot_tinneypi2} while the spread of Y0 dwarfs appears
large - the three Y0 dwarfs studied here are however very similar.  Despite
the systematic differences in parallaxes noted in Section 2.3 the mean
absolute magnitudes per spectral type derived by D\&K13 (table~S3) for the $Y,
J, H, K,$ Spitzer channel 1 \& 2 bands are consistent at the 1 sigma level
with our values.

\begin{table*}
\caption{\label{allphotometry} Apparent magnitudes from various sources and
  mean absolute magnitudes as a function of filter.}
\begin{center}
\begin{tabular}{lccccccc}
\hline\hline
Band/$\lambda_{eff}$ & $z^1$,$\sigma$ & MKO $Y^2$,$\sigma$ & MKO $J^3$,$\sigma$ & MKO $H^3$,$\sigma$ & MKO $K^2$,$\sigma$ & Spitzer ch1$^4$,$\sigma$ & Spitzer ch2$^4$,$\sigma$    \\
Target &   9535 &  10289             &  12444            & 16221              &  21900             & 35075                    &    44366                     \\
\hline                        
0410   & 22.66, 0.09   & 19.61, 0.04        & 19.34, 0.02        &  19.85, 0.03 &       19.91, 0.07        & 16.64, 0.04        & 14.17, 0.02                         \\
1738   & 22.80, 0.09   & 19.86, 0.07        & 19.55, 0.02        &  20.25, 0.03 &       20.58, 0.10        & 17.09, 0.05        & 14.47, 0.02                         \\
2056   & 23.09, 0.08   & 19.77, 0.05        & 19.20, 0.02        &  19.75, 0.02 &       20.01, 0.06        & 16.03, 0.03        & 13.92, 0.02                         \\
\hline 
$<$M$>$ &  23.57$\pm$ 0.08        & 20.45$\pm$  0.07 &           20.05$\pm$  0.07           & 20.65$\pm$  0.07         & 20.86$\pm$  0.08         & 17.24$\pm$  0.07          & 14.87$\pm$  0.07 \\
\hline\hline
Band/$\lambda_{eff}$ & WISE $W1^4$,$\sigma$ & WISE $W2^4$,$\sigma$ & WISE $W3^4,\sigma$  &  $F125W^5,\sigma $ &  $F140W^5,\sigma $ & Distance \\ 
Target &   33526               &    46028             &  115608             &         12305     &     13645  &  Modulus \\
\hline 
0410      & $>$18.170            &  14.11, 0.05 & 12.31, 0.50 &  20.00, 0.03 &  19.64, 0.02 & $0.80^{+0.14}_{-0.14}$\\
1738      & 17.71, 0.16        &  14.50, 0.04 & 12.45, 0.40 &  20.22, 0.02 & 19.92, 0.02    & $0.54^{+0.11}_{-0.10}$\\ 
2056      & 16.48, 0.08        &  13.84, 0.04 & 11.73, 0.25 &   19.81, 0.02 & 19.48, 0.02   & $0.76^{+0.10}_{-0.11}$\\
\hline 
$<$M$>$ &       17.57$\pm$  0.11          & 14.84$\pm$  0.07         & 12.71$\pm$  0.21         & 20.69$\pm$  0.07           & 20.37$\pm$  0.07 \\
\hline
\end{tabular}
\end{center}

References: 1: \citet{2013A&A...550L...2L}, 2: \citet{2015ApJ...799...37L}, 3:
This work, 4: \citet{2012ApJ...753..156K} and 5: \citet{2015ApJ...804...92S}.
The $\lambda_{eff}$ row is the simple median effective wavelength for a
convolution of the nominal filter profile and a Vega spectra, this is often
not appropriate for these objects because of the structure in the underlying
spectra but it is provided for reference. The published apparent magnitudes
are converted to absolute magnitudes using the distance modulus from the
parallaxes presented in Table~\ref{parallaxes}. The $<$M$>$ row is the
weighted mean absolute magnitude for all three targets along with the error of
the mean for each filter.
\end{table*}

%

\section{Spectroscopic Analysis and Comparison to models}

We fit the spectra from \cite{2011ApJ...743...50C} of our targets with the
atmospheric models presented in \citet{2015ApJ...804L..17T} and
\citet{2012ApJ...756..172M,2014ApJ...787...78M}. The
\citet{2015ApJ...804L..17T} model grid covers the $200 < T_{\rm eff} <
1000$\,K range with log~g values of 4.0, 4.5, and 4.8. We examined both solar
metallicity models as well as metal-poor models with [M/H]~=~-0.5 -- -0.8, and
both equilibrium and non-equilibrium models with log~$K_{\rm zz} = 6$. The
\citet{2015ApJ...804L..17T} models do not take into account clouds. The models
from \citet{2012ApJ...756..172M} are computed for $400 \leq T_{\rm eff} \leq
1300$\,K, log~g 4.0, 4.5, 5.0, and 5.5, with $2 \leq f_{\rm sed} \leq 5$. The
models from \citet{2014ApJ...787...78M} covers the $200 \leq T_{\rm eff} \leq
450$\,K range, $3.0 \leq {\rm log~g} \leq 5.0$, and $3 \leq f_{\rm sed} \leq
7$. The \citet{2014ApJ...787...78M} models assume a patchy cloud coverage with
only $50\%$ of the object being covered by clouds.

The \citet{2011ApJ...743...50C} spectra cover the wavelength range
1$-$2.4\,$\mu$m at a resolution of R~$\approx$~300 for 0410, 1.07$-$1.70\,$\mu$m
wavelength range at a low resolving power of R~$\approx$~130 for 1738, and
1.143$-$1.375\,$\mu$m and 1.431$-$1.808\,$\mu$m at a resolution or
R~$\approx$~2500 for 2056.

We derived the best-fit parameters by fitting our model grid to the observed
spectra using a standard reduced $\chi^2$ minimisation. We used two slightly
different approaches. One normalizes both target and model spectra at the peak
of the J-band (i.e. 1.26\,$\mu$m). The other approach makes use of the
parallaxes derived here. The \citet{2015ApJ...804L..17T} models provide flux
at 10\,pc (assuming a radius of 0.1\,R$_\odot$), which we have re-scaled for the
appropriate radius taken from \citet{2008ApJ...689.1327S}. We then scaled the
observed spectra to match the absolute MKO J magnitude, calculated using their
measured parallax. A similar procedure was applied to the models from
\citet{2012ApJ...756..172M} and \citet{2014ApJ...787...78M}. To assess the
quality of the best-fit parameters we adopted the approach described in
\citet{2011ApJ...743...50C}. Briefly, we generate 10,000 ``mimic spectra'' for
each object by adding Gaussian noise to the observed spectra, preserving the
signal-to-noise ratio. Then we fit this 10,000 mimics with the atmospheric
models to determine their best-fit parameters. We adopt the range of
parameters encompassed by the standard deviation of the resulting distribution
as our best-fit values.

The results are summarized in Table~\ref{teff_table}. The differences between
the best-fit parameters derived with the two methods and the different models
are small, but with the \citet{2015ApJ...804L..17T} models giving
systematically lower log~g and higher $T_{\rm eff}$ values than the
\citet{2012ApJ...756..172M,2014ApJ...787...78M} models. The reason for this 
systematic difference is beyond the scope of this contribution but it is probably due to
the use of different opacity tables, e.g. \citet{2015ApJ...804L..17T} uses
molecular line lists for ammonia from   \citet{2014MNRAS.440.1649Y} while 
\citet{2012ApJ...756..172M,2014ApJ...787...78M} uses those from
\citet{2011MNRAS.413.1828Y}.

\begin{figure*}
  \centering
    \caption*{\citet{2015ApJ...804L..17T} models.}\\
    \includegraphics[width=8.5cm]{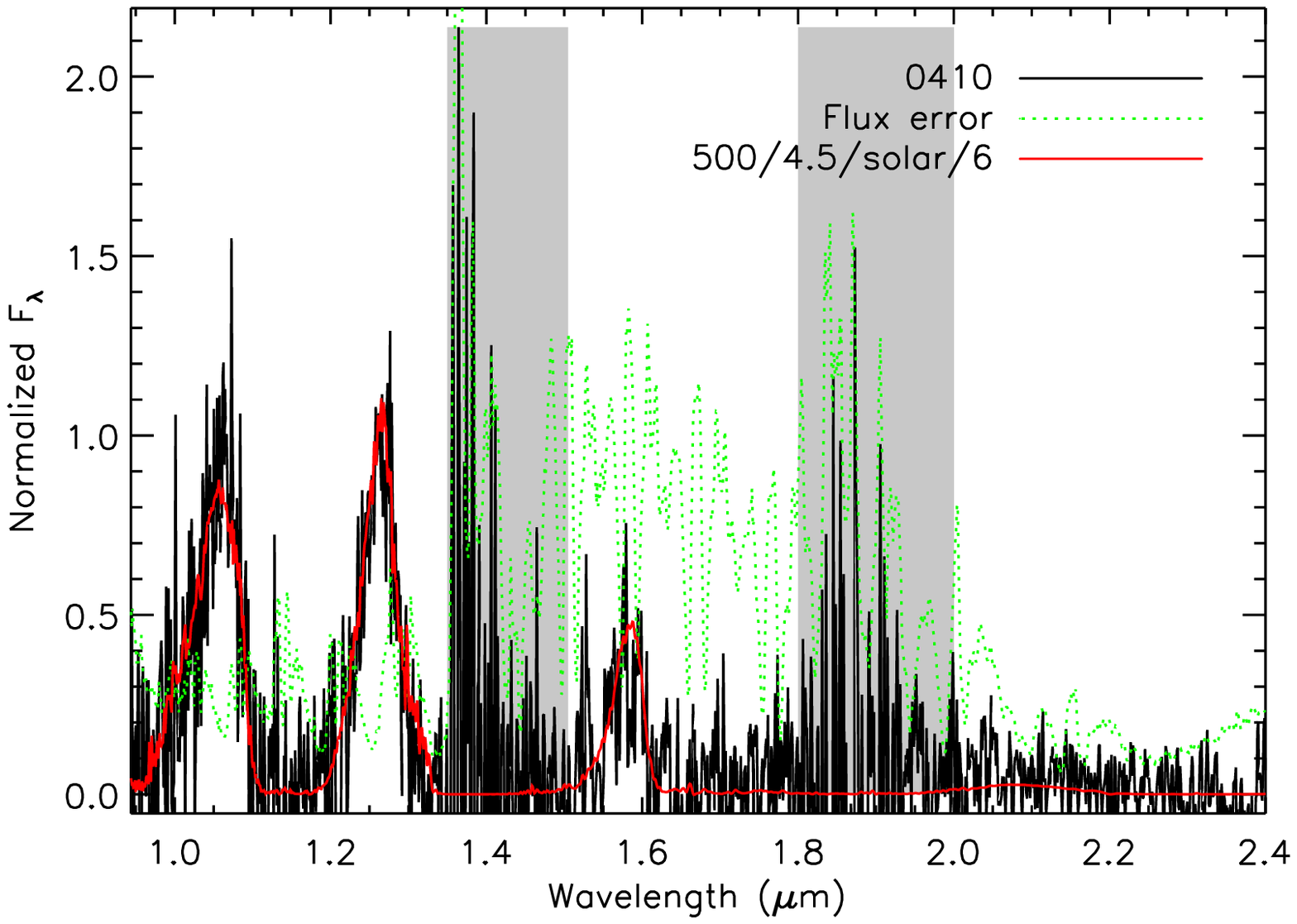}
    \includegraphics[width=8.5cm]{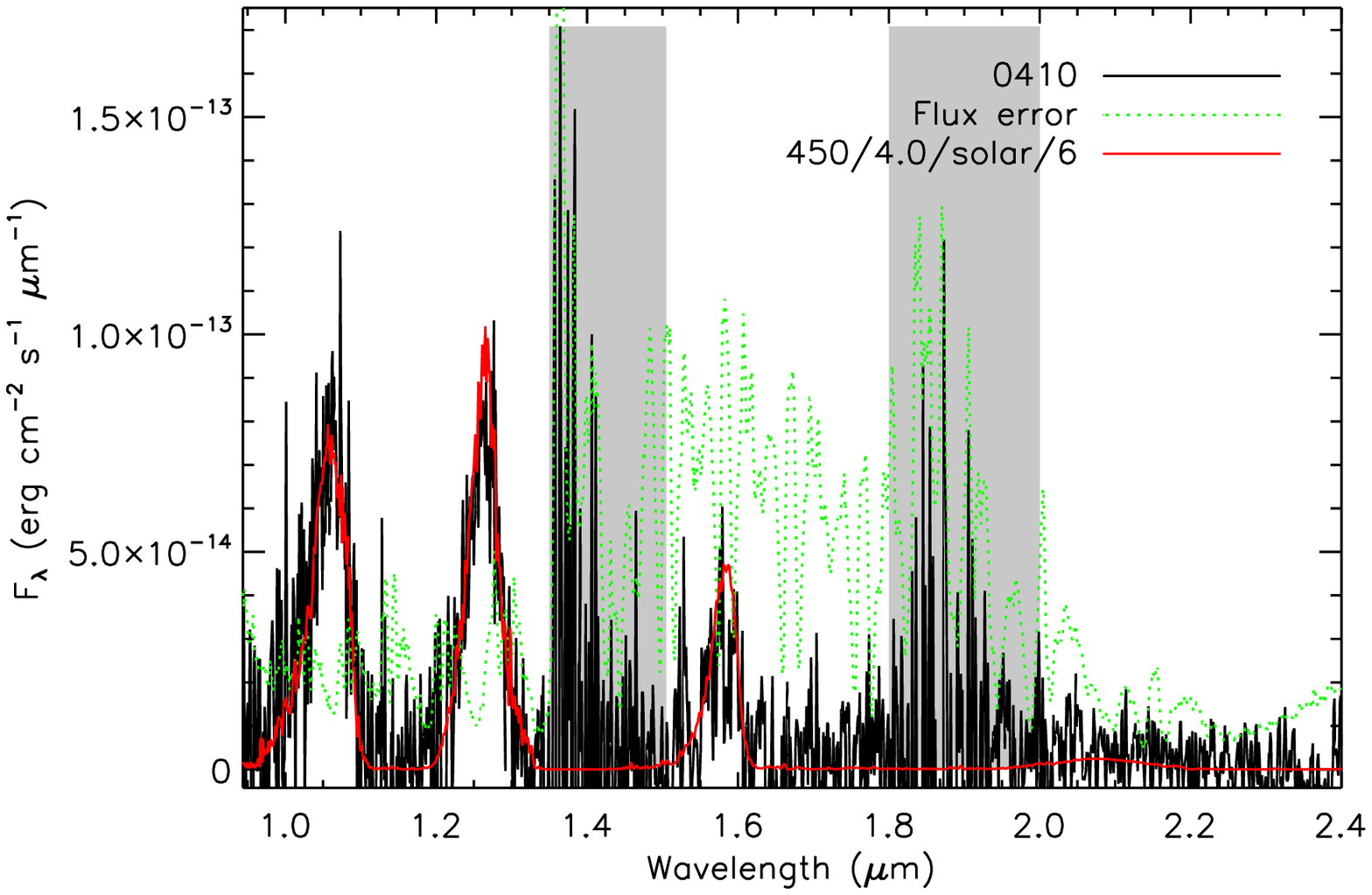}

    \caption*{\citet{2012ApJ...756..172M} models.}\\
    \includegraphics[width=8.5cm]{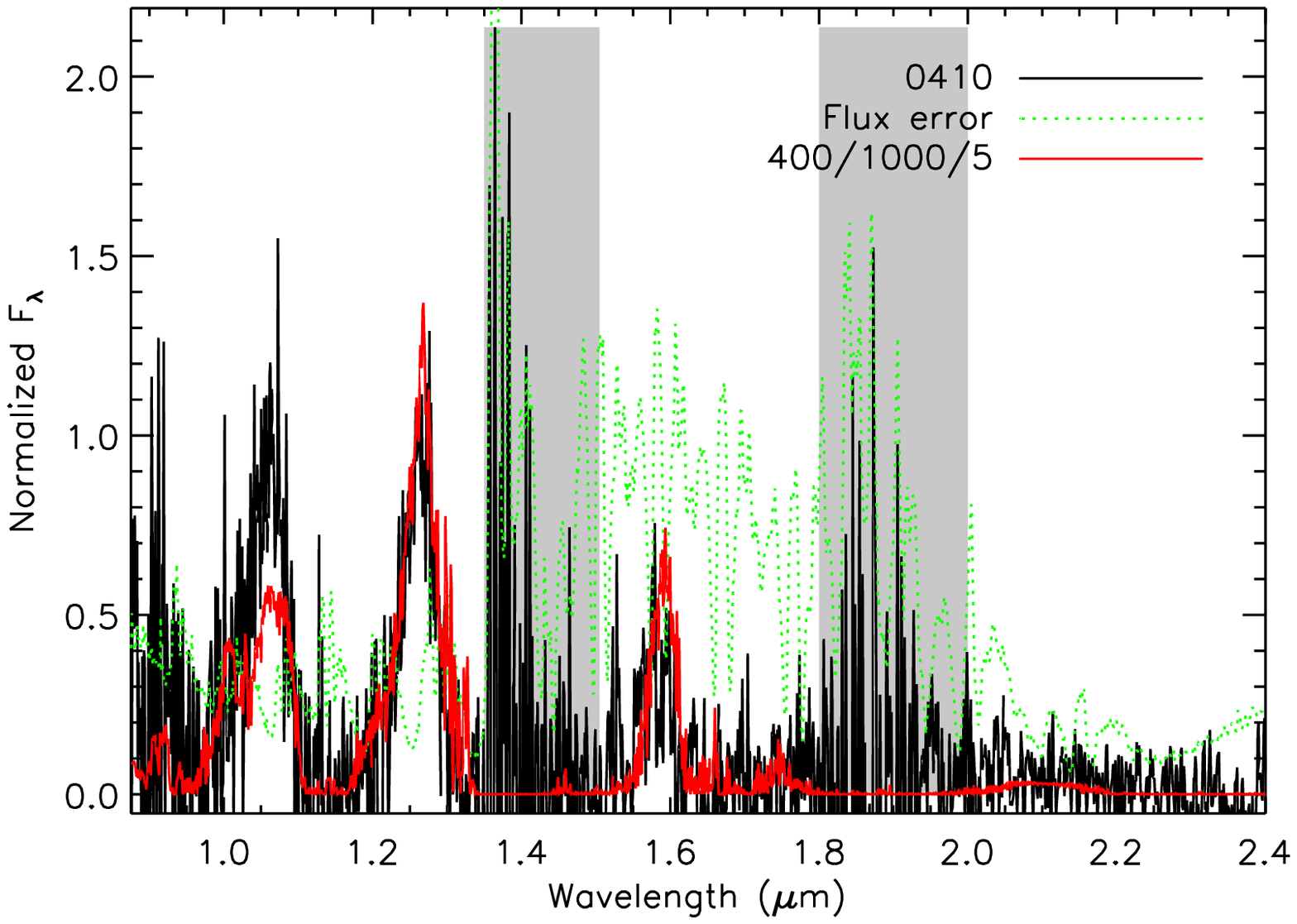}
    \includegraphics[width=8.5cm]{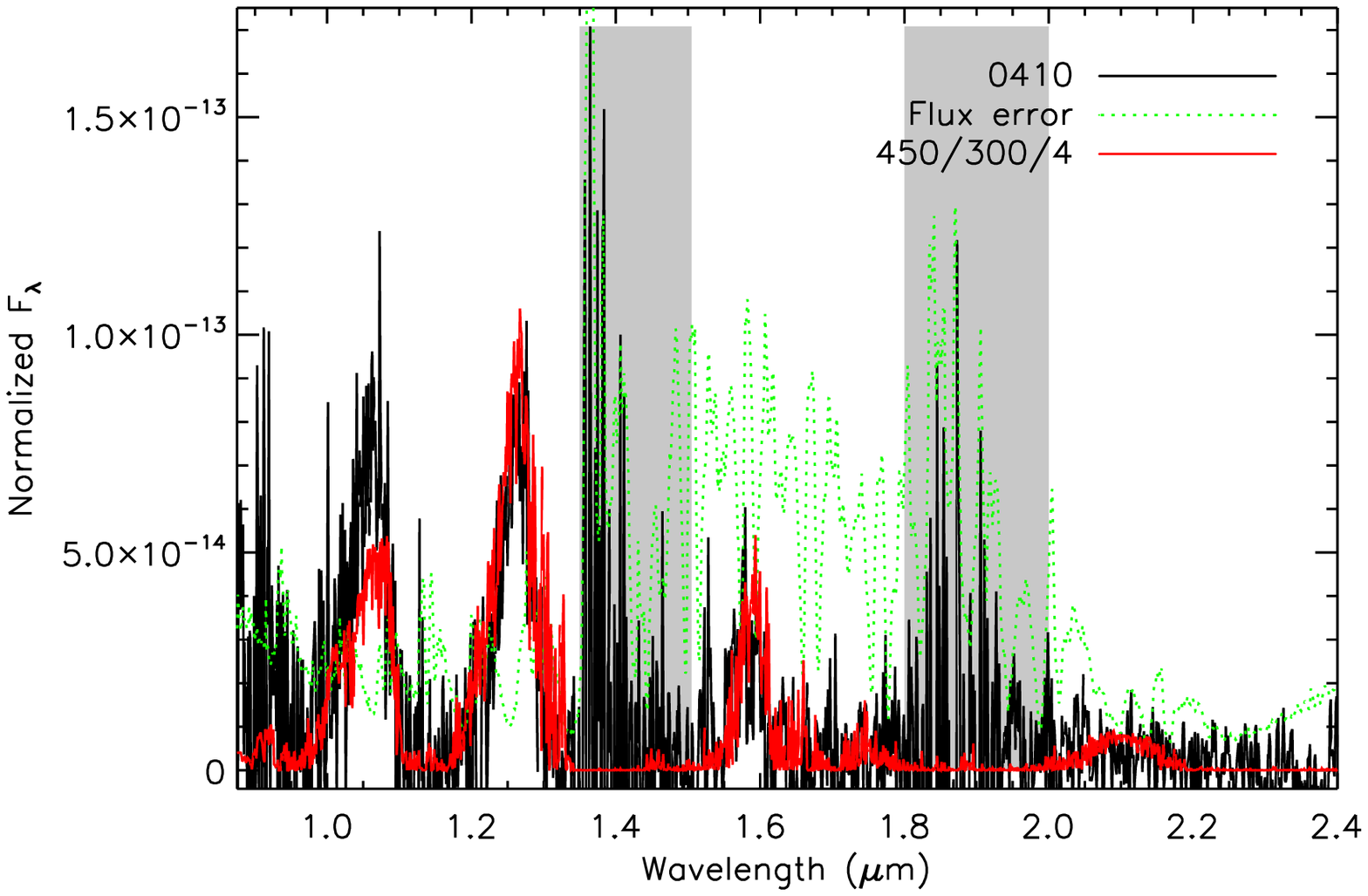}

    \caption*{\citet{2014ApJ...787...78M} models.}\\
    \includegraphics[width=8.5cm]{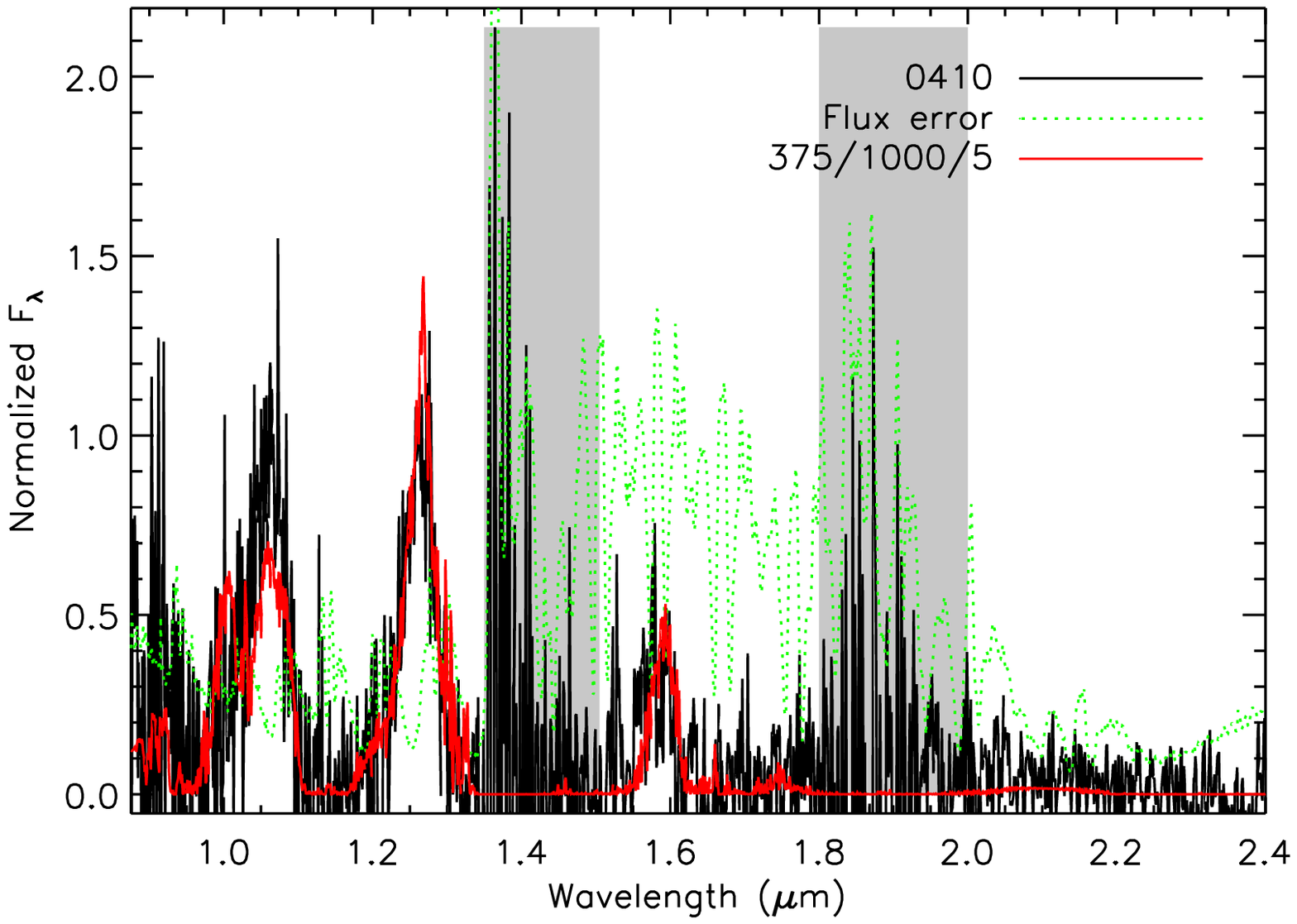}
    \includegraphics[width=8.5cm]{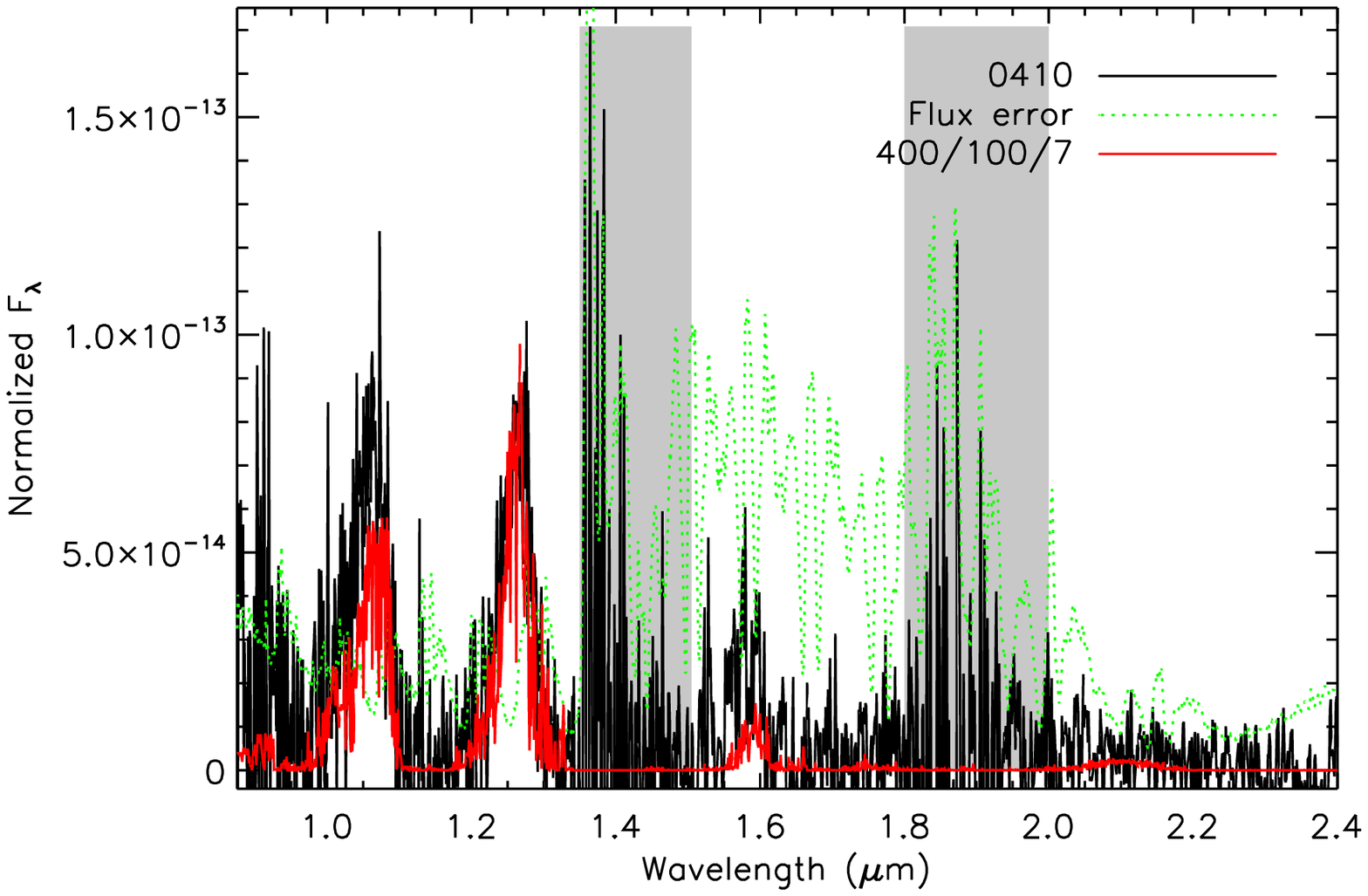}

  \caption{The best-fit model for 0410, obtained normalizing both model and
    target spectrum in the $J$ band (left panel) or scaling the observed
    spectrum to match the measured absolute $J$ magnitude (right panel). The
    spectrum of 0410 is plotted in black, its associated 1~$\sigma$
    uncertainty is plotted in green, and the best-fit model is plotted in
    red. The shaded area are regions which are heavily effected by telluric
    absorption so are excluded from the model fitting. The best-fit parameters
    are summarized in the legend, following the scheme $T_{\rm
      eff}$/log~g/[M/H]/log~$K_{\rm zz}$ for the \citet{2015ApJ...804L..17T}
    models and $T_{\rm eff}$/g/$f_{\rm sed}$ for the
    \citet{2012ApJ...756..172M} and \citet{2014ApJ...787...78M}
    models. \label{0410_model_fit}}
\end{figure*}

For 0410 the best-fit parameters when normalizing the spectra are $T_{\rm eff}
\sim 450$~K, log~g~$\sim$~4.5, and solar metallicity. The only notable
exception are the \citet{2014ApJ...787...78M} models, that would predict a
lower $T_{\rm eff}$ of 350$-$400~K. When using the measured parallax we obtain
a slightly lower $T_{\rm eff}$. The spectrum of 0410 and the best fit models
are plotted in Fig.~\ref{0410_model_fit}. We note that 500\,K is a higher
temperature than previously found for Y0s
\citep[e.g. ][]{2012ApJ...753..156K,2013ApJ...763..130L, 2014ApJ...783...68B},
while the value obtained using the parallax is more in line with the published
estimates.
  
\begin{figure*}
  \centering
    \caption*{\citet{2015ApJ...804L..17T} models.}\\
    \includegraphics[width=8.5cm]{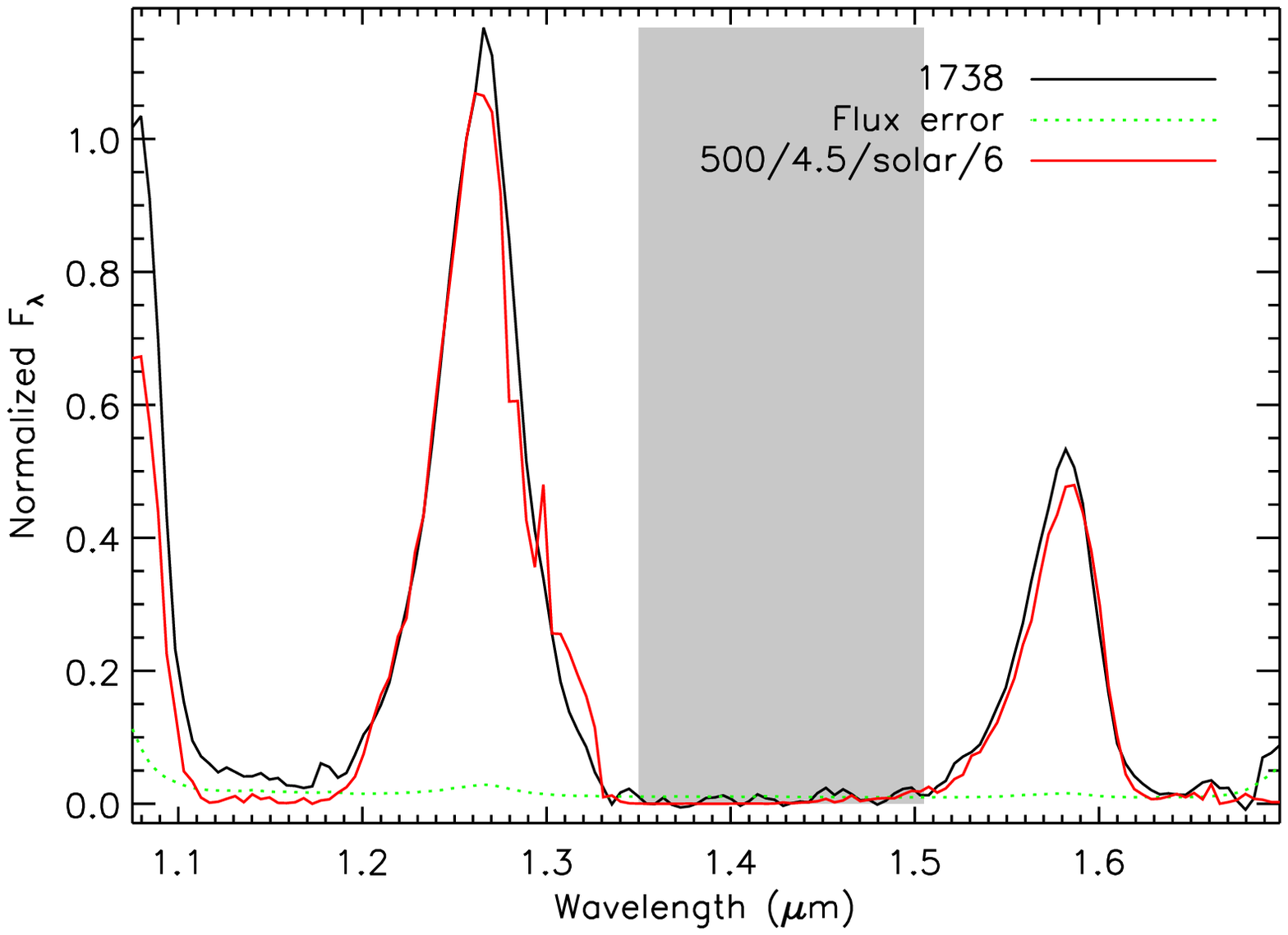}
    \includegraphics[width=8.5cm]{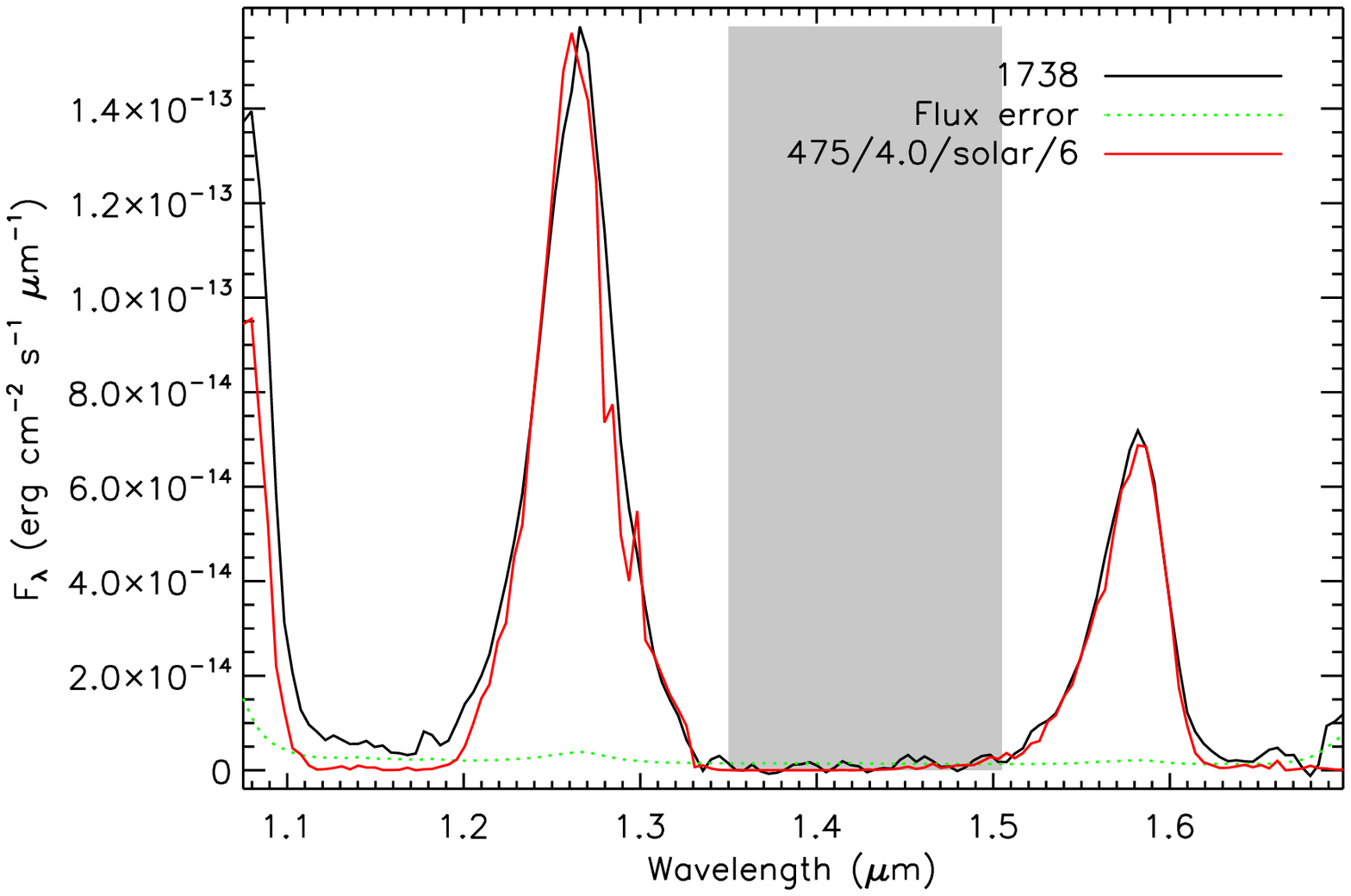}

    \caption*{\citet{2012ApJ...756..172M} models.}\\
    \includegraphics[width=8.5cm]{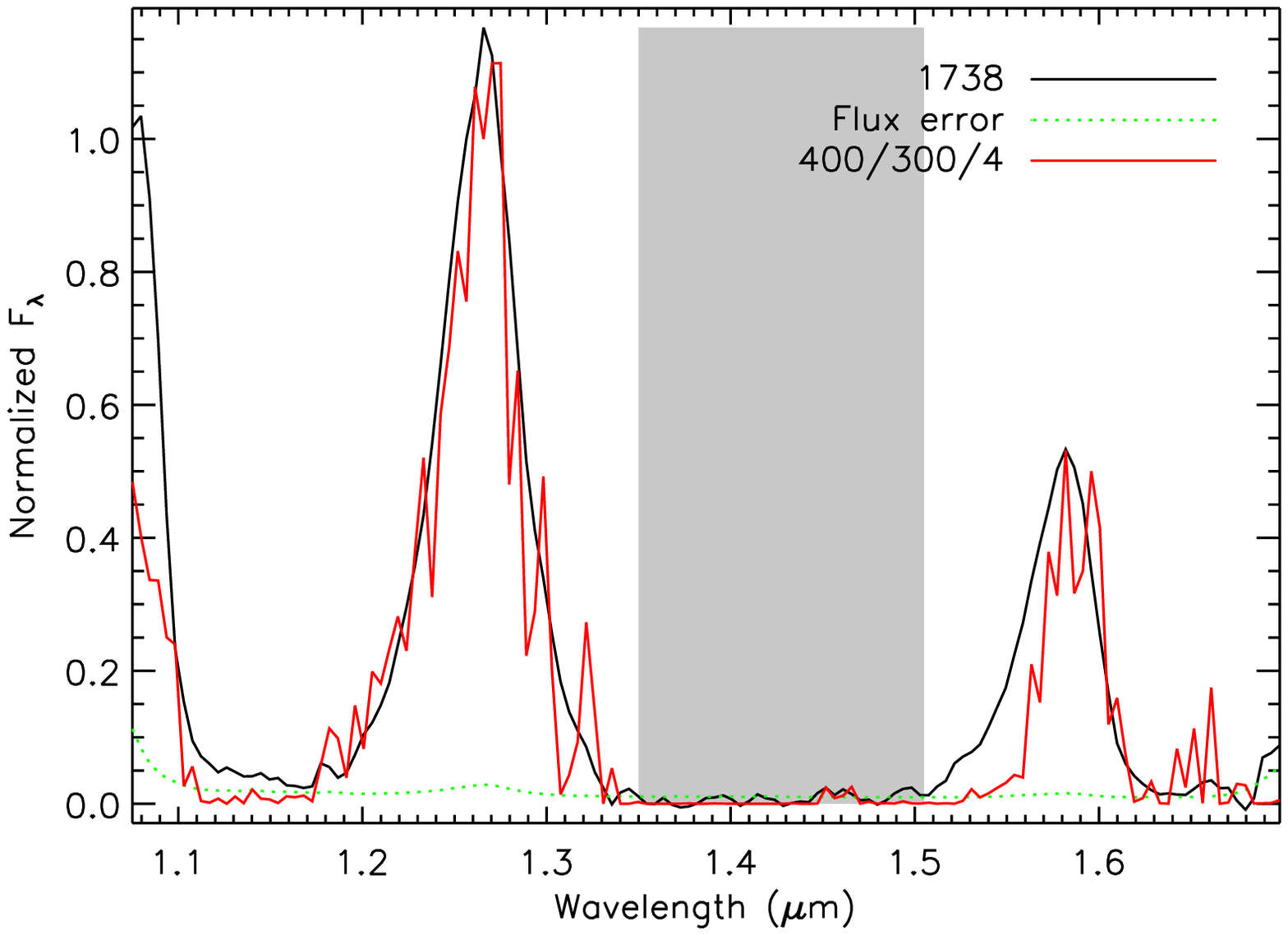}
    \includegraphics[width=8.5cm]{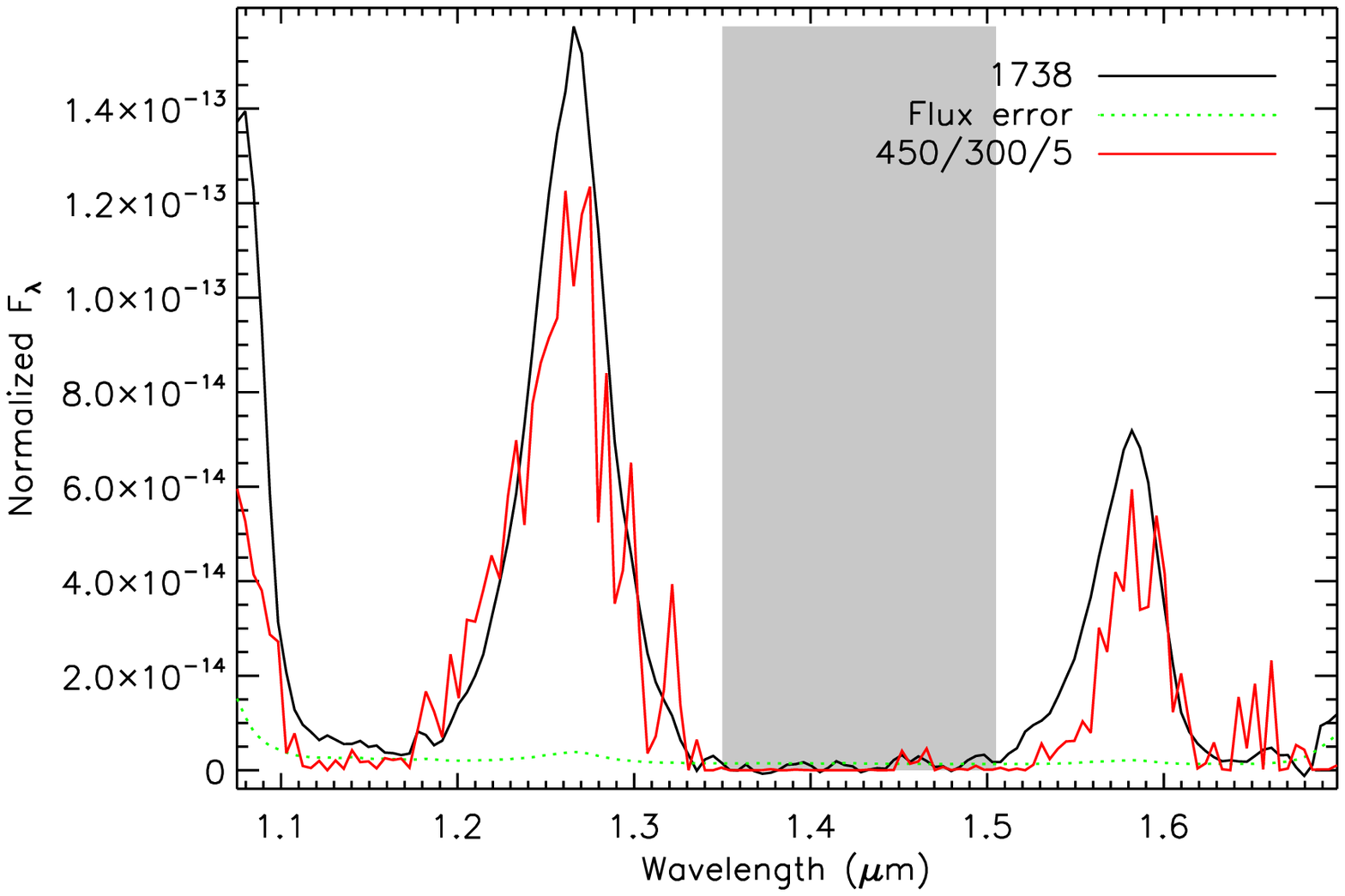}

    \caption*{\citet{2014ApJ...787...78M} models.}\\
    \includegraphics[width=8.5cm]{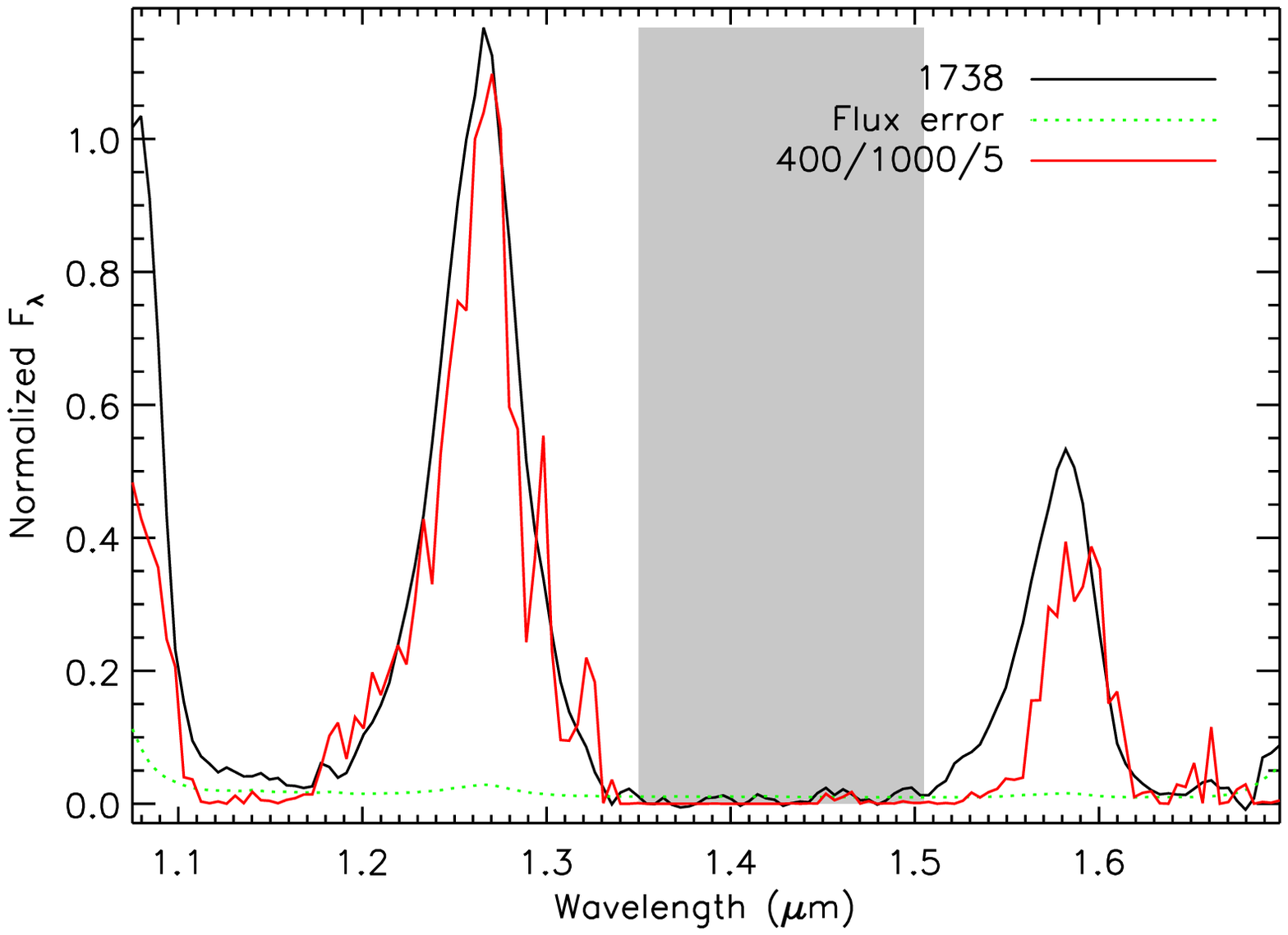}
    \includegraphics[width=8.5cm]{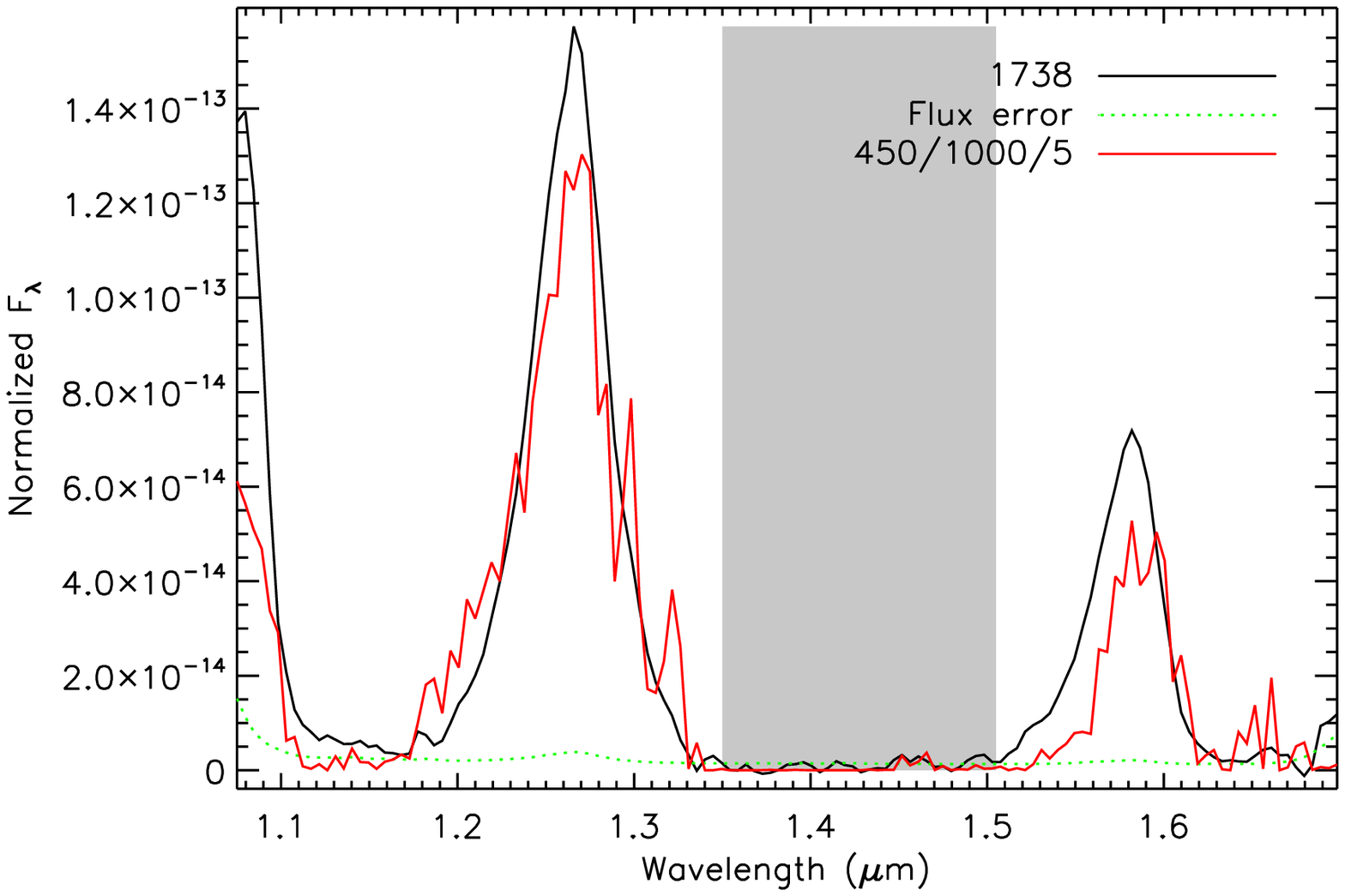}

  \caption{Same as Fig.~\ref{0410_model_fit} but for 1738. \label{1738_model_fit}} 
\end{figure*}

For 1738 the best-fit parameters given by the various models tend to be more
discordant. When normalizing the spectra the \citet{2015ApJ...804L..17T}
models give $T_{\rm eff} = 500$\,K and log~g~$=~4.5$ (with solar metallicity
and log~$K_{\rm zz} = 6$). The other models predict a lower $T_{\rm eff}$ of 400\,K
with a higher log~g of $\approx~4.75$. Using our measured parallax mitigates
the discrepancy, with the \citet{2012ApJ...756..172M,2014ApJ...787...78M}
models returning best-fit $T_{\rm eff} = 450$\,K and log~g~$=~4.5-5.0$, closer
to the \citeauthor{2015ApJ...804L..17T} values. The spectrum of 1738 and the
best fit models are plotted in Fig.~\ref{1738_model_fit}.

\begin{figure*}
  \centering
    \caption*{\citet{2015ApJ...804L..17T} models.}\\
    \includegraphics[width=8.5cm]{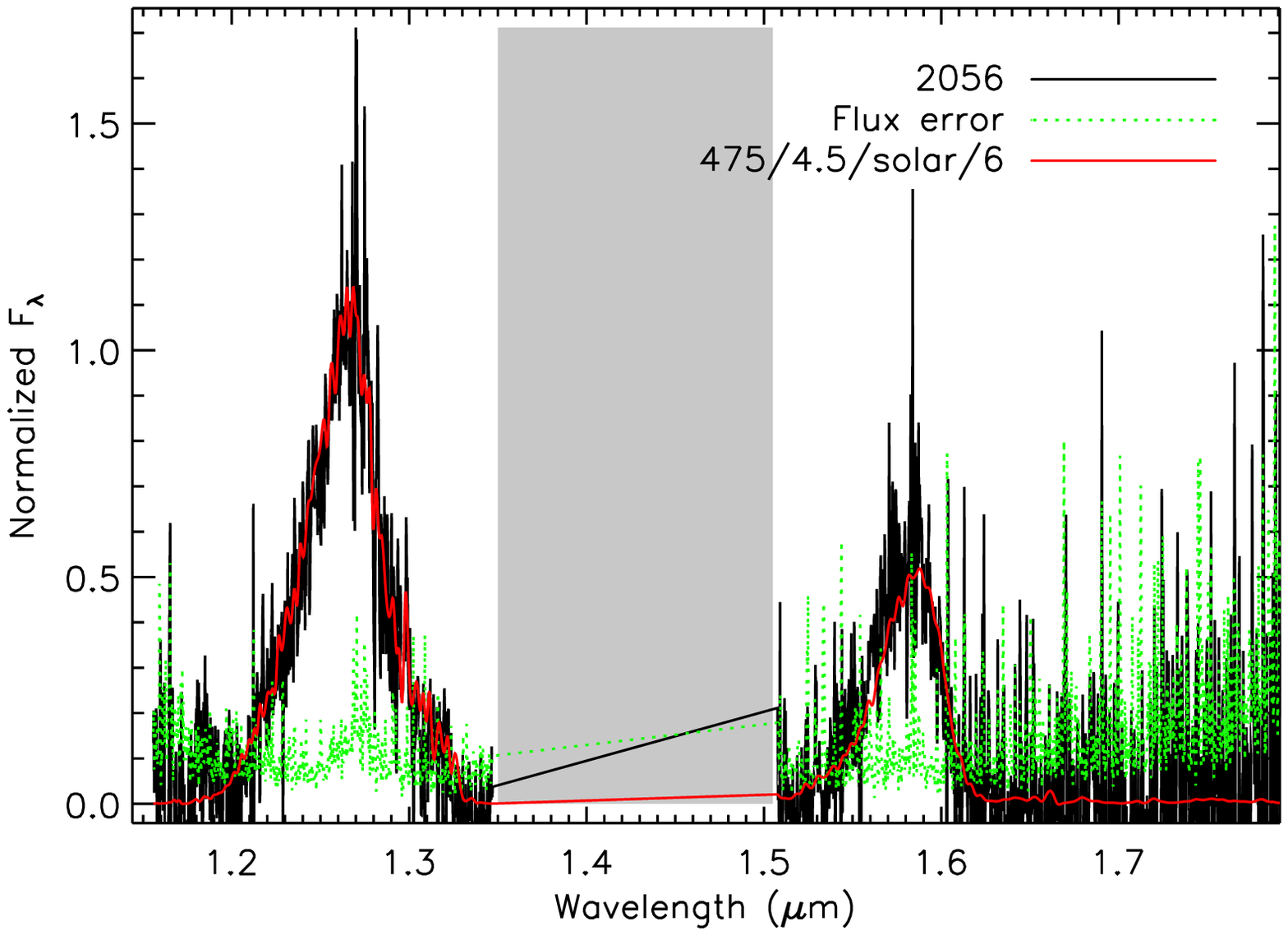}
    \includegraphics[width=8.5cm]{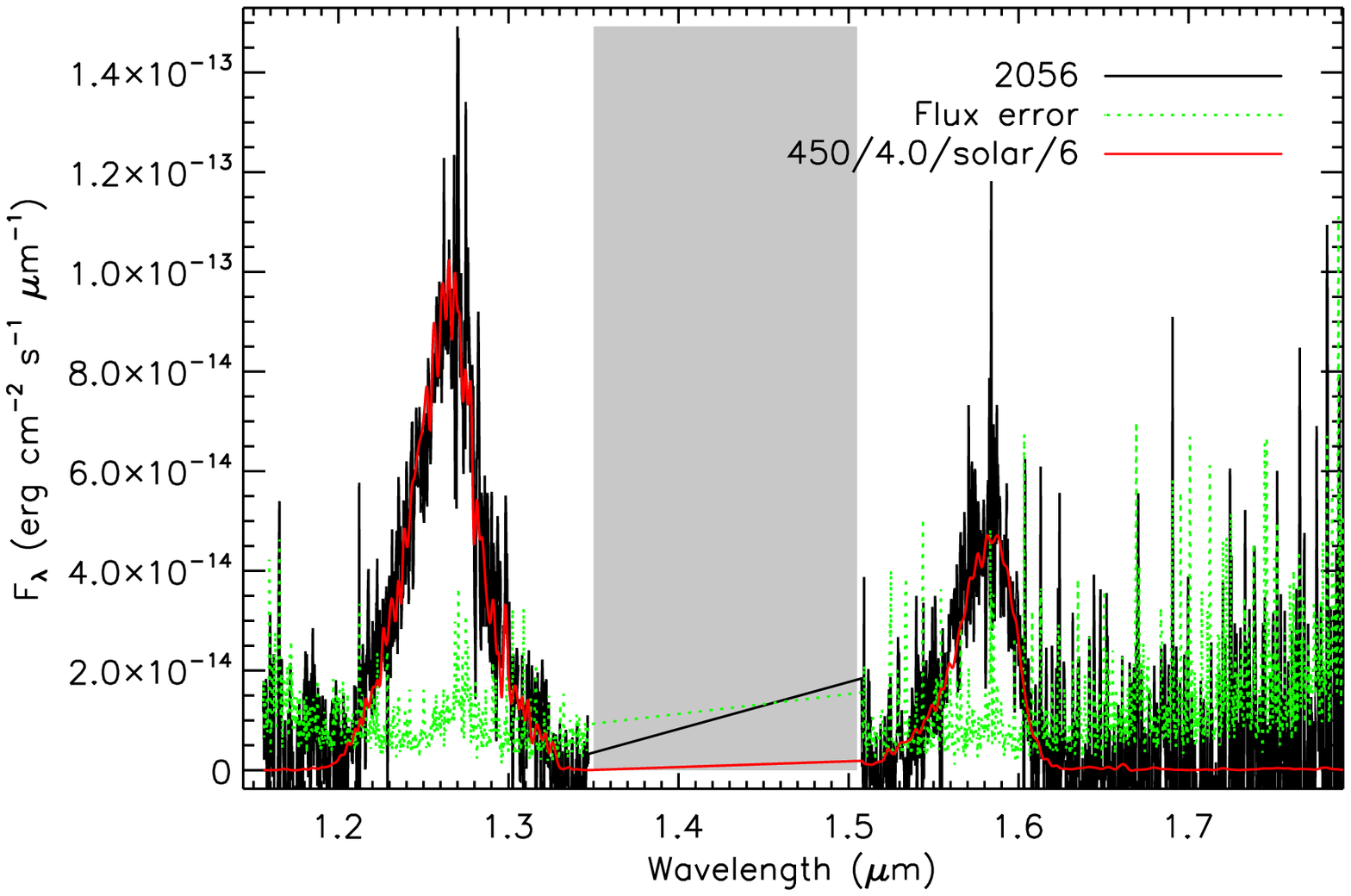}

    \caption*{\citet{2012ApJ...756..172M} models.}\\
    \includegraphics[width=8.5cm]{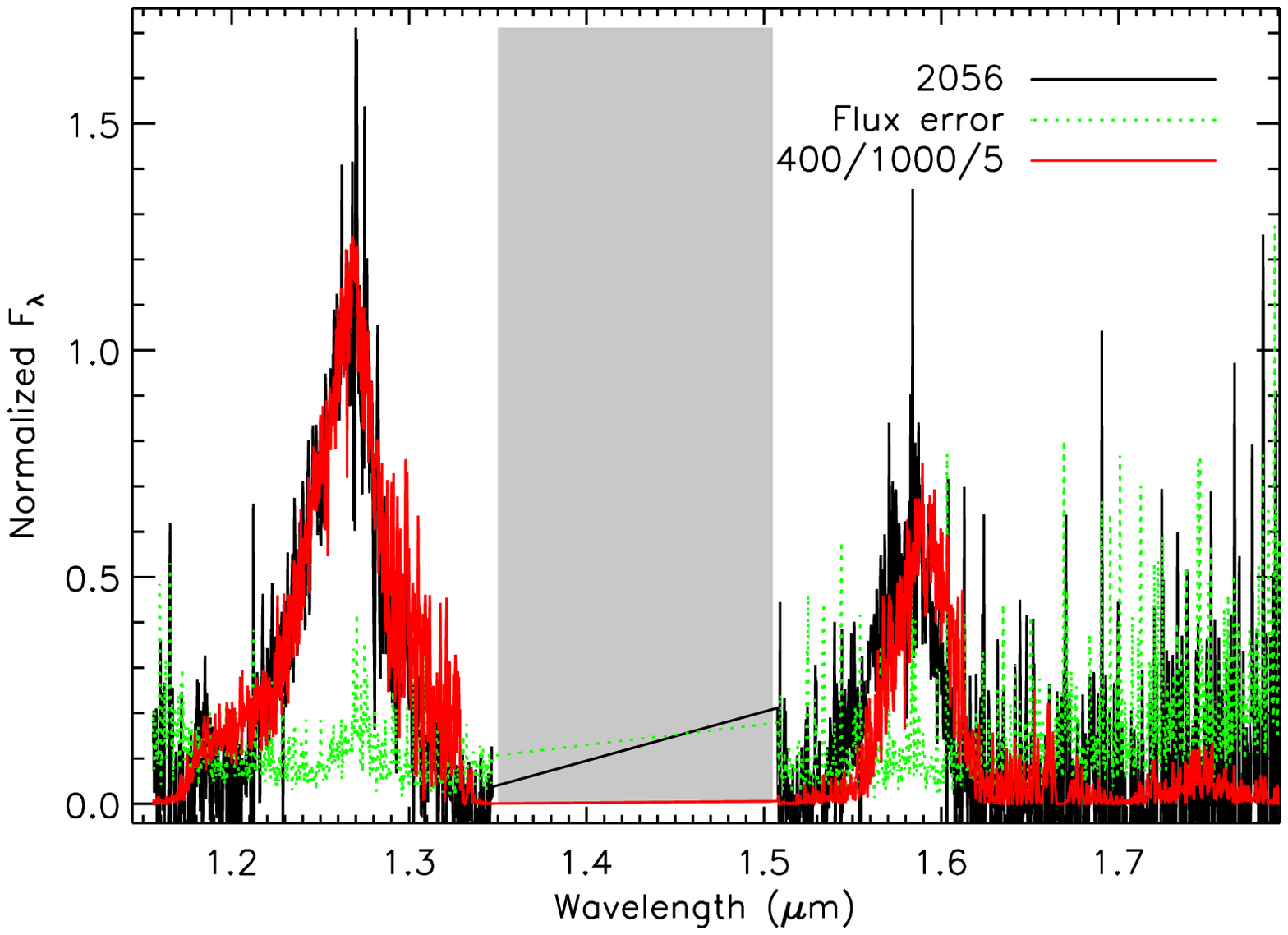}
    \includegraphics[width=8.5cm]{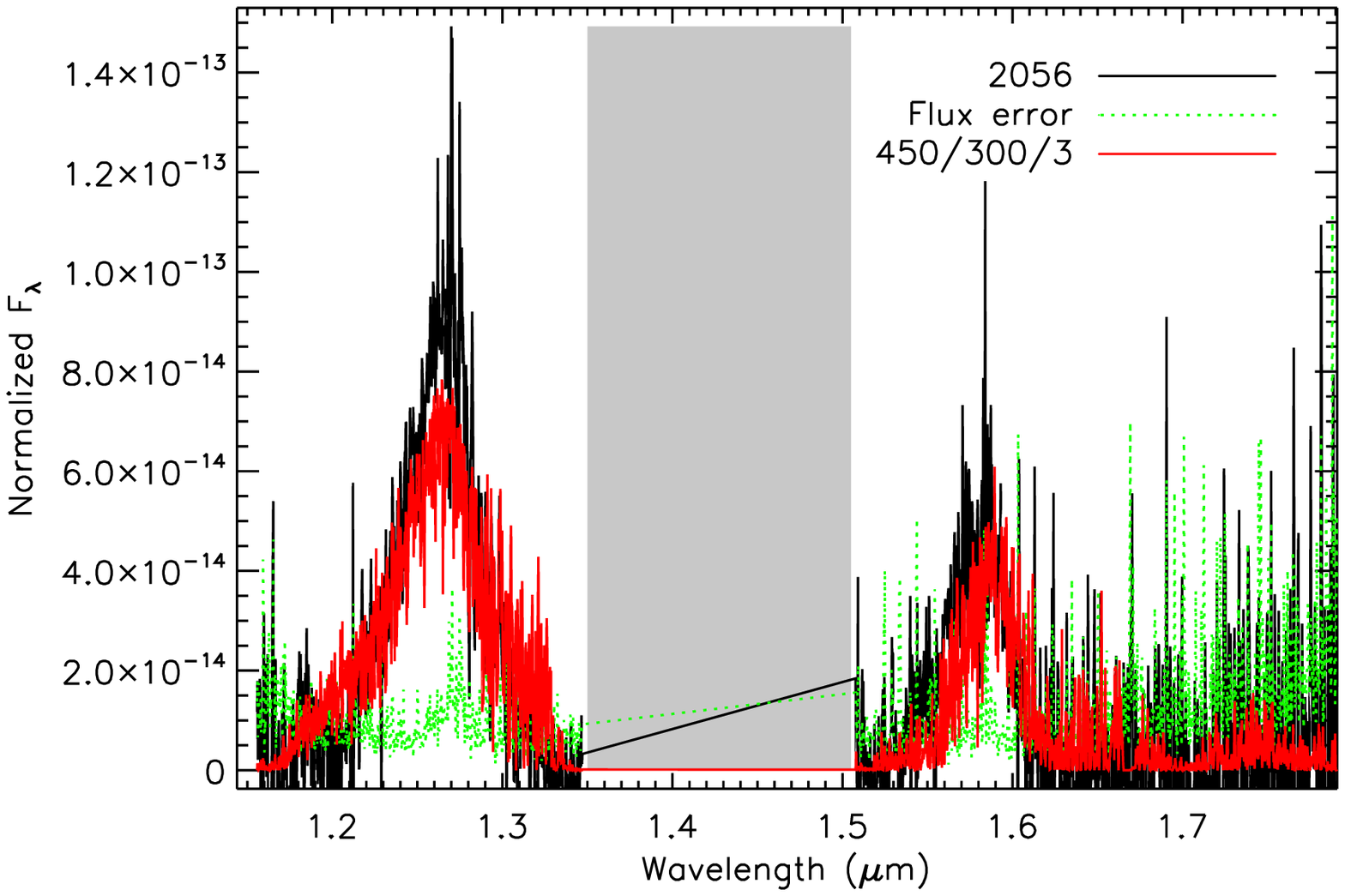}

    \caption*{\citet{2014ApJ...787...78M} models.}\\
    \includegraphics[width=8.5cm]{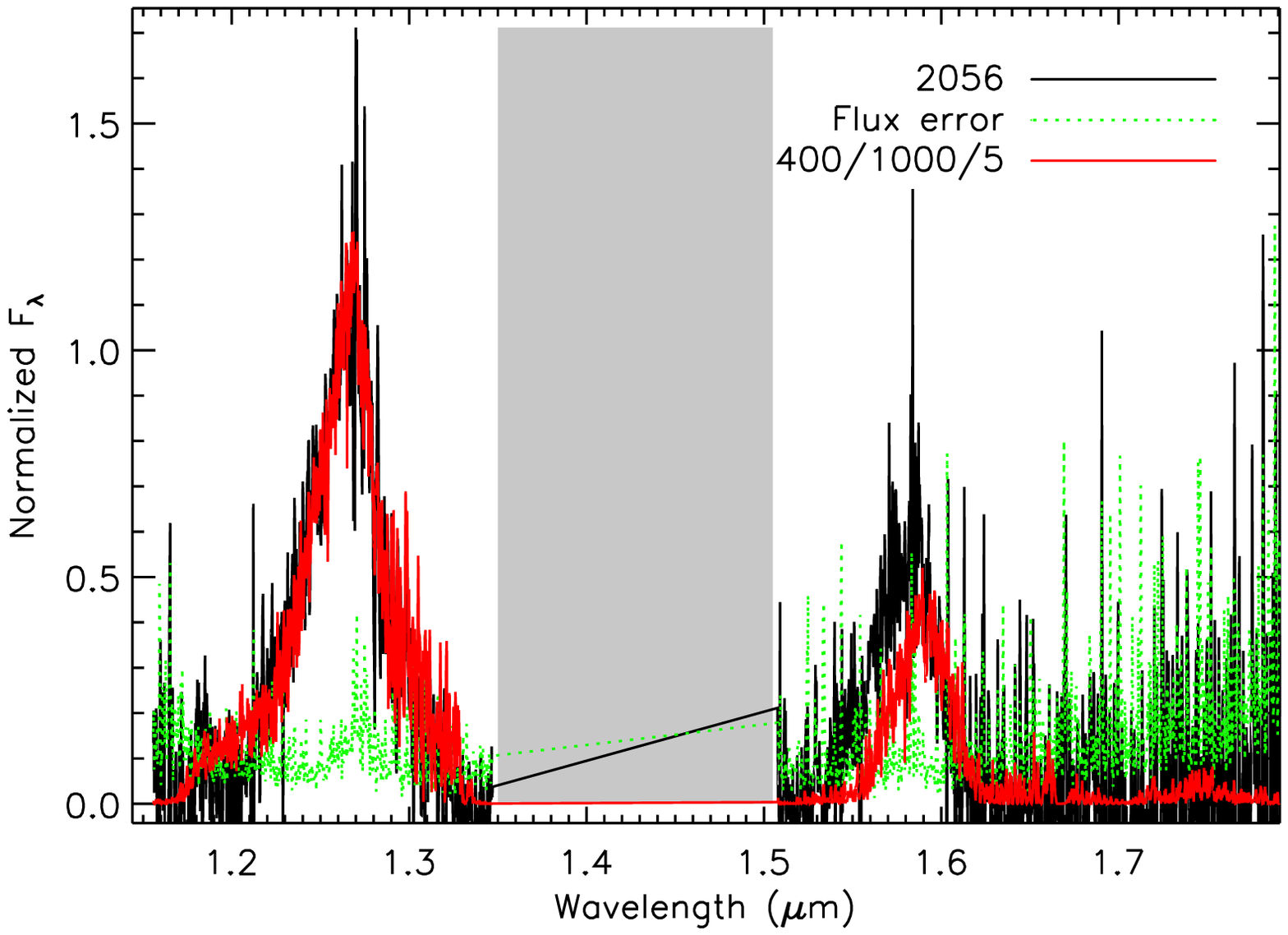}
    \includegraphics[width=8.5cm]{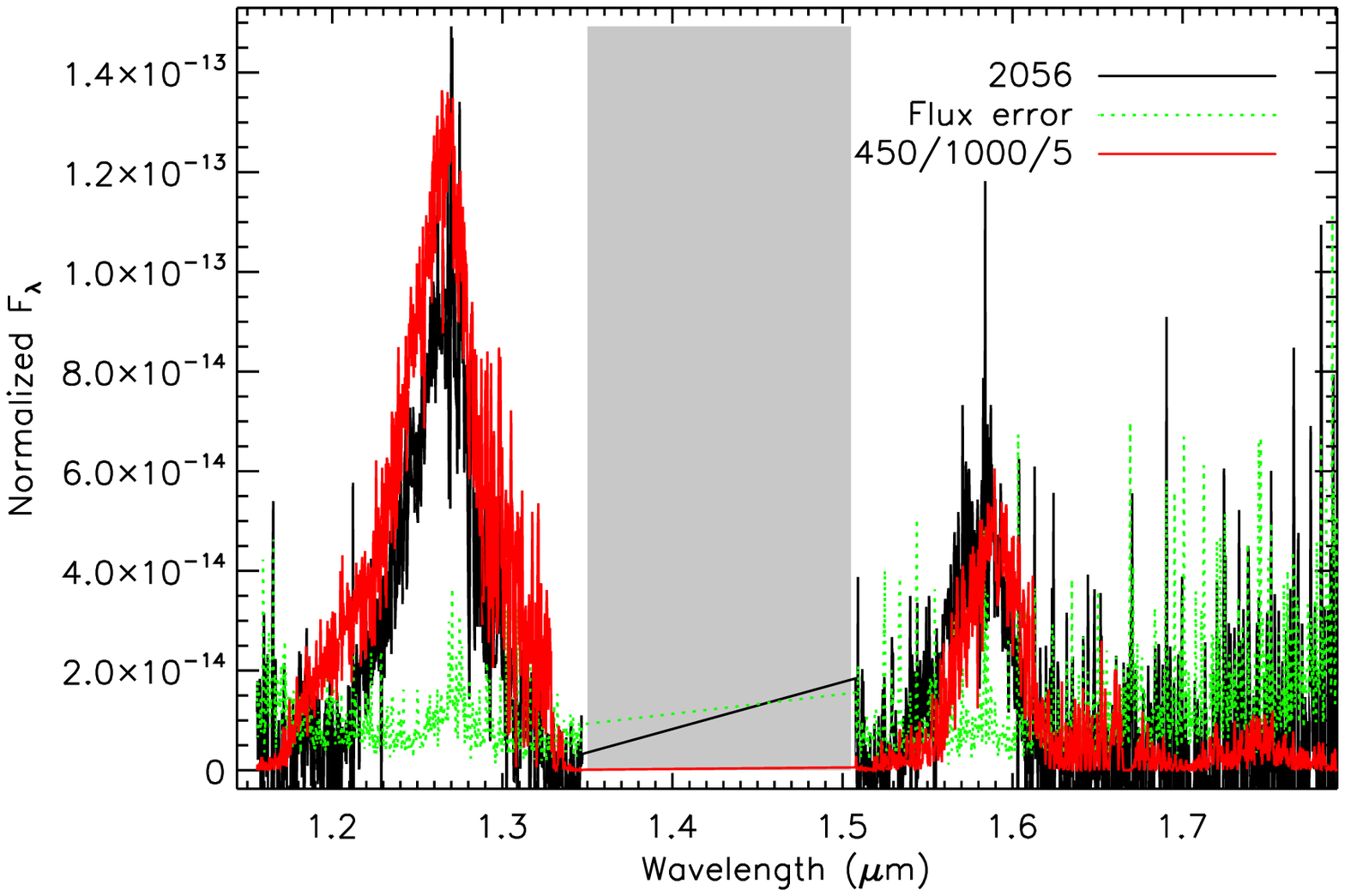}
  \caption{Same as Fig.~\ref{0410_model_fit} but for 2056. \label{2056_model_fit}} 
\end{figure*}

For 2056 we have similar situation, with some discrepancy between the best-fit
parameters derived from the different models when normalizing the spectra. The
\citet{2015ApJ...804L..17T} models predict slightly higher $T_{\rm eff}$
compared to the other models (450$-$500\,K vs. 375$-$450\,K), and also in this
case the discrepancy is removed when using the measured parallax. The spectrum
of 2056 and the best fit models are plotted in Fig.~\ref{2056_model_fit}.

\begin{table*}
  \begin{tabular}{c|ccc|ccc|ccc|ccc}
          & \multicolumn{9}{|c|}{This paper} & \multicolumn{3}{c}{\citet{2011ApJ...743...50C}} \\
    Models & \multicolumn{3}{c|}{\citet{2015ApJ...804L..17T}} & \multicolumn{3}{c|}{\citet{2012ApJ...756..172M}} & \multicolumn{3}{c|}{\citet{2014ApJ...787...78M}} & \multicolumn{3}{c}{\citet{2008ApJ...689.1327S}} \\
    \hline
    Short & $T_{\rm eff}$ & log~g & log          & $T_{\rm eff}$ & log~g & $f_{\rm sed}$ & $T_{\rm eff}$ & log~g & $f_{\rm sed}$ & $T_{\rm eff}$ & log~g & log \\       
    Name  & (K)           &       & $K_{\rm zz}$ & (K)           &       &               & (K)           &       &               & (K)           &       &  $K_{\rm zz}$ \\
    \hline
    \hline
    \multirow{2}{*}{0410}  & 500       & 4.0$-$4.5 & 6 & 400$-$450 & 4.0$-$5.0 & 4$-$5 & 350$-$400 & 4.5$-$5.0 & 5     & 450 & 3.75 &  0 \\   
                           & 450       & 4.0$-$4.5 & 6 & 450       & 4.0$-$4.5 & 3$-$5 & 375$-$400 & 4.0       & 5$-$7 &     &      &  \\     
    \hline
    \multirow{2}{*}{1738}  & 475$-$500 & 4.0$-$4.5 & 6 & 400       & 4.5$-$5.0 & 4$-$5 & 400       & 5.0       & 5     & 350 & 4.75 &  4 \\   
                           & 475       & 4.0       & 6 & 450       & 4.5       & 5     & 450       & 5.0       & 5     &     &      &  \\  
    \hline   
    \multirow{2}{*}{2056}  & 450$-$500 & 4.0$-$4.5 & 6 & 400$-$450 & 4.5$-$5.0 & 4$-$5 & 375$-$400 & 5.0       & 5     & 350 & 4.75 &  4 \\   
                           & 450       & 4.0       & 6 & 450       & 4.5       & 3$-$4 & 400$-$450 & 5.0       & 5     &     &      &  \\     
    \hline
    \hline
  \end{tabular}
  \caption{The best-fit parameters obtained with our model fitting, compared
    with those obtained by \citet{2011ApJ...743...50C}. For each target, the
    first line presents the parameters obtained normalizing both model and
    target spectrum in the $J$ band, while the second line presents the
    parameters obtained scaling the observed spectrum to match the measured
    absolute $J$ magnitude. The metallicity, [M/H], was a fitted parameter for
    the \citet{2015ApJ...804L..17T} models, but always came out to be solar
    for all sets of parameters. \label{teff_table}}

\end{table*}

\section{Discussion}

We have found parallaxes with relative errors better than 7\% using
observations from just one telescope and detector combination. The three Y0
objects studied here have absolute magnitudes that are consistent to within
the observational errors and we find mean absolute magnitudes for various
pass-bands. A comparison to other parallaxes determinations shows our values
to be consistent with those of Beichman et al (2014) but indicate the values
in Dupuy \& Kraus (2013) are underestimated.  While Dupuy \& Kraus published
only relative parallaxes the observed difference is too large to be due to a
required correction to an absolute value.

This difference must lie in the reduction procedures or in some
systematic bias in the observational material. Parallaxes of objects this
faint will remain the domain of relative small field programs for the near
future but with wealth of objects being published by Gaia a gold standard will
be produced which will allow small field programs to calibrate instruments and
refine procedures to the sub-mas level. Also, while Gaia will not observe
directly any objects cooler than late T dwarfs it will indirectly detect cool
objects in binary systems that will serve as direct comparisons in absolute
magnitude space to those measured with small field programs. The ability to
detect and remove systematic errors from the observations and reductions along
with a direct comparison sample will lead to more robust and consistent small
field results that will only aid the astrophysical interpretation.

As shown in Section 3 our multi-epoch observations sample the near-infrared
brightness evolution of the three targets over a baseline of about 4 years.
Simple linear fits to the $J$ measurements show small and marginally
statistically significant long-term slopes in two of our targets and indicate
a possible slope in the third target (Table~\ref{spitzermagstable}).
Although the existence of long-term monotonic variations is tentative, this
possibility is intriguing and in the following we briefly explore their
possible nature.

Our measurements represent the first long-term precision photometry of Y
dwarfs and, as such, the first probes of long-term atmospheric evolution in
these objects.  Recent precision near-infrared (Spitzer) studies of Y-dwarfs
detected high-amplitude ($\sim$3\%--15\%) rotational modulations in three
targets: WISEP J140518.40+553421.5 \citep{2016ApJ...823..152C}; WISE
J173835.52+273258.9 \citep{2016ApJ...830..141L} and WISE
J085510.83--071442.5 \citep{2016ApJ...832...58E}.  These modulations are
periodic on the 6 hour timescales, and consistent with rotational modulation
that are found to be common in L, L/T, and T-type brown dwarfs by extensive
and sensitive space-based surveys
\citep{2014ApJ...782...77B,2015ApJ...799..154M}. Detailed analysis of
photometric variability in L/T dwarfs indicated variations of cloud properties
\citep[e.g.][]{2009ApJ...701.1534A,2012ApJ...750..105R}, which was verified as
correlated temperature and cloud thickness variations by high precision
time-resolved spectroscopy \citep{2013ApJ...768..121A}. In contrast, for mid-L
type brown dwarfs modulations in the condensate cloud properties failed to
reproduce the observed gray variations and indicate a high-level haze layer
\citep{2016ApJ...826....8Y}. The observed modulations in Y-dwarfs are probably
emerging due to the rotational modulations introduced by heterogeneous KCl and
Na$_2$S clouds \citep[][]{2016arXiv160707888L}.

Unlike the above studies, our observations provide a sensitive probe of
Y-dwarfs over {\em very long timescales}, i.e., over $10^4$ rotations. Our
data hints on the possible existence of {\em monotonic} changes, which is not
consistent with stochastic cloud evolution (occurring over dynamical
timescales, about a rotational period).  Although the full interpretation of
the variations is beyond the scope of this work, we embark here on a
speculative discussion of the importance of the photometric variations.  The
changes, if real, are probably occurring due to a long-term monotonic
evolution of the clouds, driven by a process that acts on a timescale much
longer than the dynamical timescales. We note that chemical disequilibrium
could drive such slow changes if the kinetic timescales for one or more
important processes are very long.  An example of such slow, chemical
disequilibrium-driven cloud evolution is given by the spectacular 2012 Saturn
Storm \citep[][]{2013Icar..226..402S}. During this event water vapor-rich air
was dredged up from the deep interior of Saturn to its upper, cold and very
dry atmosphere. Water and other volatiles froze out in the upper atmosphere
and formed clouds that were optically thick at optical and infrared
wavelengths. Over the course of the following six months the clouds
encompassed Saturn's northern hemisphere and eventually dispersed, with ice
crystals likely settling to the deeper interior, leaving the upper atmosphere
dry again.  We speculate here that qualitatively similar events may drive
long-timescale monotonic brightness evolution in Y-dwarfs.

We have compared the combination of spectra and parallaxes to the atmospheric
models of Tremblin et al. (2015) and Morley et al. (2012, 2014). We find the
best physical parameters are consistent between the three objects with solar
metallicity, temperatures between 450-475\,K, and a log~g of 4.0-4.5.  A general
consideration that arises from the model fitting is that, if we do not employ
the measured parallax, the models alone would lead to overestimated log~g,
i.e. we would essentially be overestimating the mass of our targets. We note
that for \citet{2015ApJ...804L..17T} all best fit models are non-equilibrium
models, stressing the importance of mixing in such low-temperature
atmospheres. All \citet{2015ApJ...804L..17T} best-fit models are
solar-metallicity models, however given our coarse metallicity grid, this
result is less conclusive. All Y dwarfs have best-fit log~g~$\geq$~4.0,
suggestive of ``old'', evolved objects (age~$>$~1~Gyr), in agreement with
their kinematics. If we compare the parameters derived here with those
presented in \citet{2011ApJ...743...50C}, we note that we either get a
significantly higher $T_{\rm eff}$ ($\sim 100 - 150$\,K higher, for 1738 and
2056), or a significantly higher log~g (0.75~dex, for 0410). While comparing
the results from different model grids is a dangerous exercise, (given the
disparate underlying assumptions, the different parameter space covered, and
the heterogeneous steps of the grid), it is however remarkable to see such a
large discrepancy in the predicted atmospheric parameters. Finally, we stress
that this analysis is based on near-infrared spectra only, while Y dwarfs emit
most of their energy at longer wavelengths. Any systematic issue with models
in the near-infrared range would therefore lead to incorrect atmospheric
parameters.

\newpage

\section{Acknowledgments}
We thank the anonymous referee for comments that improved the clarity of this
contribution; Isabelle Baraffe and Gilles Chabrier for useful discussions on
the model aspects addressed here; Luca Rizzi, Tom Kerr, Watson Varricatt and
Andy Adamson for scheduling help; and Mike Cushing who promptly provided the
spectra of these three targets. This work has made use of the Cambridge
Astronomy Survey Unit software and the WFCAM Science Archive thanks to Mike
Irwin and Mike Read for their support. The United Kingdom Infrared Telescope
was operated by the Joint Astronomy Centre on behalf of the Science and
Technology Facilities Council of the U.K., it is currently operated by the
University of Arizona. RLS's research was supported by a Henri Chr\'etien
International Research Grant administered by the American Astronomical Society
and a Visiting Professorship with the Leverhulme Trust (VP1-2015-063). SKL's research is
supported by the Gemini Observatory, which is operated by the Association of
Universities for Research in Astronomy, Inc., on behalf of the international
Gemini partnership of Argentina, Australia, Brazil, Canada, Chile and the
United States of America.  FM/HRAJ/DJP acknowledge support from the UK's
Science and Technology Facilities Council grant number ST/M001008/1. The
collaboration was supported by the Marie Curie 7th European Community
Framework Programme grant n.247593 {\it Interpretation and Parameterisation of
  Extremely Red COOL dwarfs} (IPERCOOL) International Research Staff Exchange
Scheme.

\bibliographystyle{mnras}
\bibliography{refs}

\end{document}